\documentclass[%
 reprint,
 amsmath,amssymb,
 aps,
 prd,
]{revtex4-2}

\usepackage{graphicx}
\usepackage{dcolumn}
\usepackage{bm}
\usepackage{xcolor}
\usepackage{float}
\usepackage{enumitem}
\usepackage{orcidlink}

\DeclareMathOperator{\cR}{\mathcal{R}}
\DeclareMathOperator{\LieXi}{\mathcal{L}_{\xi}}

\begin{document}

\preprint{APS/123-QED}

\title{On the Asymptotic Evolution of Bulk-viscous, Spherically Symmetric Spacetimes}

\author{Balázs Endre Szigeti\orcidlink{0000-0002-8028-962X}}\email{szigeti.balazs@wigner.hun-ren.hu}
\affiliation{HUN-REN Wigner Research Centre for Physics, P.O. Box 49, H-1525 Budapest, Hungary}
\affiliation{Eötvös Loránd University, Institute of Computer Science, 11/C Pázmány Péter Stny,}
\affiliation{Eötvös Loránd University, Institute of Physics, 11/A Pázmány Péter Stny, H-1117 Budapest Hungary} 

\author{Imre Ferenc Barna\orcidlink{0000-0001-6206-3910}}%
 \email{barna.imre@wigner.hun-ren.hu}
\affiliation{HUN-REN Wigner Research Centre for Physics, P.O. Box 49, H-1525 Budapest, Hungary}

\author{Gergely Gábor Barnaföldi\orcidlink{0000-0001-9223-6480}}
 \email{barnafoldi.gergely@wigner.hun-ren.hu}
\affiliation{HUN-REN Wigner Research Centre for Physics, P.O. Box 49, H-1525 Budapest, Hungary} 

\date{\today}

\begin{abstract}
The scale-free nature of gravitational interaction in both Newtonian gravity and the general theory of relativity gives rise to the concept of self-similarity, where solutions are scale invariant. As a result of this property, the governing partial differential equations are greatly simplified and can be transformed into ordinary ones. These solutions function as attractors, characterizing the asymptotic dynamics of more general solutions. There exist situations in which self-similarity is only partially realized, giving rise to kinematic self-similar solutions. Our study provides a systematic classification of kinematic self-similar solutions corresponding to the most general spherically symmetric space-time in the presence of bulk viscous flows.
\end{abstract}

\maketitle

\section{Introduction}

In general relativity, self-similarity can be realized in two different ways. On one hand, it could appear as a property of spacetime and, on the other hand, as a property of matter fields~\cite{https://doi.org/10.48550/arxiv.gr-qc/0405113}. Such self-similar behavior often reflects the asymptotic evolution of space-time, offering insight into its long-term dynamics. From the mathematical cosmology aspect, the main interest has so far been focused on self-similar solutions with a perfect fluid stress-energy tensor. In this context, extending the analysis beyond the idealized perfect fluid case naturally leads to considering more general matter sources. Among these, viscous fluids provide an important generalization, as dissipative processes can play a significant role in the dynamics of cosmological models.
 
 The description of viscosity and irreversible processes in general relativity has a long history. C. H. Eckart proposed the first model in 1940~\cite{PhysRev.58.919}. Later, W. A. Lindblom and L. Hiscock~\cite{PhysRevD.31.725} demonstrated that Eckart's proposal yields theories that are unstable and non-causal. Despite recognizing these problems since 1985, Eckart's first-order theory is widely used in cosmology. In our analysis, we have used the Landau\,--\,Lifschitz\,--\,Eckart flow, in spite of the existence of a second-order theory -- proposed by Muller, Israel and Sterwart \cite{Israel1979-mis}-- which addresses the issues arising in Eckart's formulation~\cite{Freistuhler2014-lo}. K. Misner proposed the first widely recognized viscous cosmological model in Ref.~\cite{Misner1986}. He introduced viscosity in his cosmological model from a particle physics standpoint.

The great interest in viscous cosmology models emerges from one of the most fundamental concepts in hydrodynamics, that the ideal description is only an approximation of real-world systems. Moreover, as it was articulated by Maartens ('95) \cite{Maartens_1995}:

\emph{“The conventional theory of the evolution of the universe includes several dissipative processes, as it must if the current large value of the entropy per baryon is to be accounted for. (...) important to develop a robust model of dissipative cosmological processes in general, so that one can analyze the overall dynamics of dissipation without getting lost in the details of particular complex processes.”} In light of this argument, it is clear why many new viscous models have gained prominence in the scientific discourse of recent years~\cite{doi:10.1142/S0218271817300245, Brevik2015-sh}.

Self-similarity implies that the spatial distribution of the characteristics of motion remains identical to itself at all times throughout the process. This is a well-known concept in classical physics; it was introduced by Gottfried Guderley in 1942 and used extensively in various fields of physics \cite{Guderley1942}.  Also, self-similar \emph{ansatz} represent solutions to degenerate problems in which all dimensional parameters entering the initial and boundary conditions vanish or become infinite~\cite{Barenblatt1996}. Since then, self-similarity has become a standard tool in modern theoretical and applied physics, providing valuable insight into scale-invariant structures in systems governed by nonlinear dynamics~\cite{Csorgo2004-ch,Barna2014}. 

Self-similarity holds great importance in Newtonian gravitational theory, making it a useful tool for describing various systems. However, its generalization to general relativity (GR) is not quite straightforward, due to its covariant nature. It was first defined by Cahill and Taub \cite{Cahill1971} and has been extensively studied for various space-time symmetries~\cite{Gad2021, Sharif2006}. It has a wide range of applications in mathematical cosmology~\cite{Eardley1974, Carr1974}, the gravitational collapse of black holes~\cite{Gundlach2007}, and the study of naked singularities~\cite{PhysRevD.42.1068}.

In Sec.~\ref{sec::Theory}, we present the basic concepts of self-similarity in general relativity and show its application to general, non-static spherically symmetric space-time. We establish the relevant equation system to find all kinematic self-similar solutions in the presence of a specific viscous fluid --- using the generalization of the framework proposed by Cahill and Taub~\cite{Cahill1971}. In Sec.~\ref{sec:results} we present the relevant ordinary differential equation systems arising from the self-similar \emph{ansatz}, derive their analytic solutions, and discuss their classification in the presence of bulk viscosity terms. The basic notation and conventions used in this paper can be found in Appendix~\ref{Sec::Append1}.
\section{\label{sec::Theory} Kinematic self-similarity of general spherical symmetric space-time}
Originally, self-similarity in general relativity can be defined via the existence of a special type of Killing vector field called homothetic vector fields (HVF)~\cite{Stephani2003, Cahill1971}. The defining relation for HVFs is,
\begin{equation} \label{eq::homothety}
\mathcal{L}_{\boldsymbol{\xi}}g_{\mu \nu} = 2 \delta g_{\mu \nu},
\end{equation}
where $\delta$ is a constant, the $g_{\mu \nu}$ is the metric tensor and $\mathcal{L}_{\boldsymbol{\xi}}$ is the Lie-differentiation along $\boldsymbol{\xi}$. In general, it can be proved that $\delta$ can be set to unity by rescaling the $\boldsymbol{\xi}$ vector field. Note that if $\delta$ equals zero, the homothetic vector fields become the well-known Killing vector fields. If the Eq.~\eqref{eq::homothety} holds for a $\boldsymbol{\xi}$ homothetic vector field, and the Riemann-tensor, hence the Ricci-tensor and therefore the Einstein-tensor, meet the
\begin{equation} \label{eq::Riemann}
    \mathcal{L}_{\boldsymbol{\xi}} R^{\mu}_{\nu \kappa \lambda} = 0 \  \Rightarrow \ \mathcal{L}_{\boldsymbol{\xi}} R_{\mu \nu} = 0 \ \Rightarrow \ \mathcal{L}_{\boldsymbol{\xi}} G_{\mu \nu} = 0, 
\end{equation}
identity, respectively~\cite{Hawking1973}. Contrarily, if $\boldsymbol{\xi}$ satisfies the so-called collineation equations~\eqref{eq::Riemann}, it does not mean that it is a homothetic vector field. Non-vacuum systems are described via
\begin{equation} \label{eq::Einstein}
    G_{\mu \nu} = \kappa T_{\mu \nu}.
\end{equation}
Einstein field equations (EFEs) are satisfied if the perfect fluid. The energy-momentum tensor must meet the collinearity condition of
\begin{equation} \label{eq::LieDerivativeT}
    \mathcal{L}_{\boldsymbol{\xi}} T_{\mu \nu} = 0.
\end{equation}
The $T_{\mu \nu}$ can be given by the standard form
\begin{equation}
    T_{\mu \nu} = (\rho + p) u_{\mu} u_{\nu} + p g_{\mu \nu}, 
\end{equation}
where $\rho$ is the energy density, $p$ denotes the pressure, and the $u_{\mu}$ is the four-velocity.
The original article by Cahill and Taub showed that only the linear barotropic equation of state $(p = w_0\rho)$ is compatible with this homothety condition for perfect fluid~\cite{Cahill1971}. Comprehensive analysis of homothetic vector field on perfect fluid was mase by Eardly~\cite{Eardley1974}. In his work, Eardly identified "physical" self-similarity with the "geometric" homothetic condition. However, if the source is not a perfect fluid, spacetime symmetries need not be inherited by the matter. Hence, a homothety remains a purely geometric property rather than a manifestation of self-similarity~\cite{Carot1994,Coley1991-ah}. Exact self-similar solutions have been obtained across a variety of spacetime symmetries: for plane-symmetric spacetimes~\cite{Foglizzo1993, Shikin1979}, hyperbolically symmetric spacetimes~\cite{Chao1981}, Weyl spacetimes~\cite{Godfrey1972}, and spherically symmetric spacetimes~\cite{PhysRevD.62.044022,Carr1999}, highlighting the broad applicability of self-similarity in relativistic contexts. Such models provide insight into the critical phenomena observed at the threshold of black hole~\cite{Choptuik1993, Gundlach2007}, as well as primordial black hole formation~\cite{osti_6856773,Lin1976}.

\subsection{Kinematic self-similar solutions}

\begin{figure}
    \centering
    \includegraphics[width=1.0\linewidth]{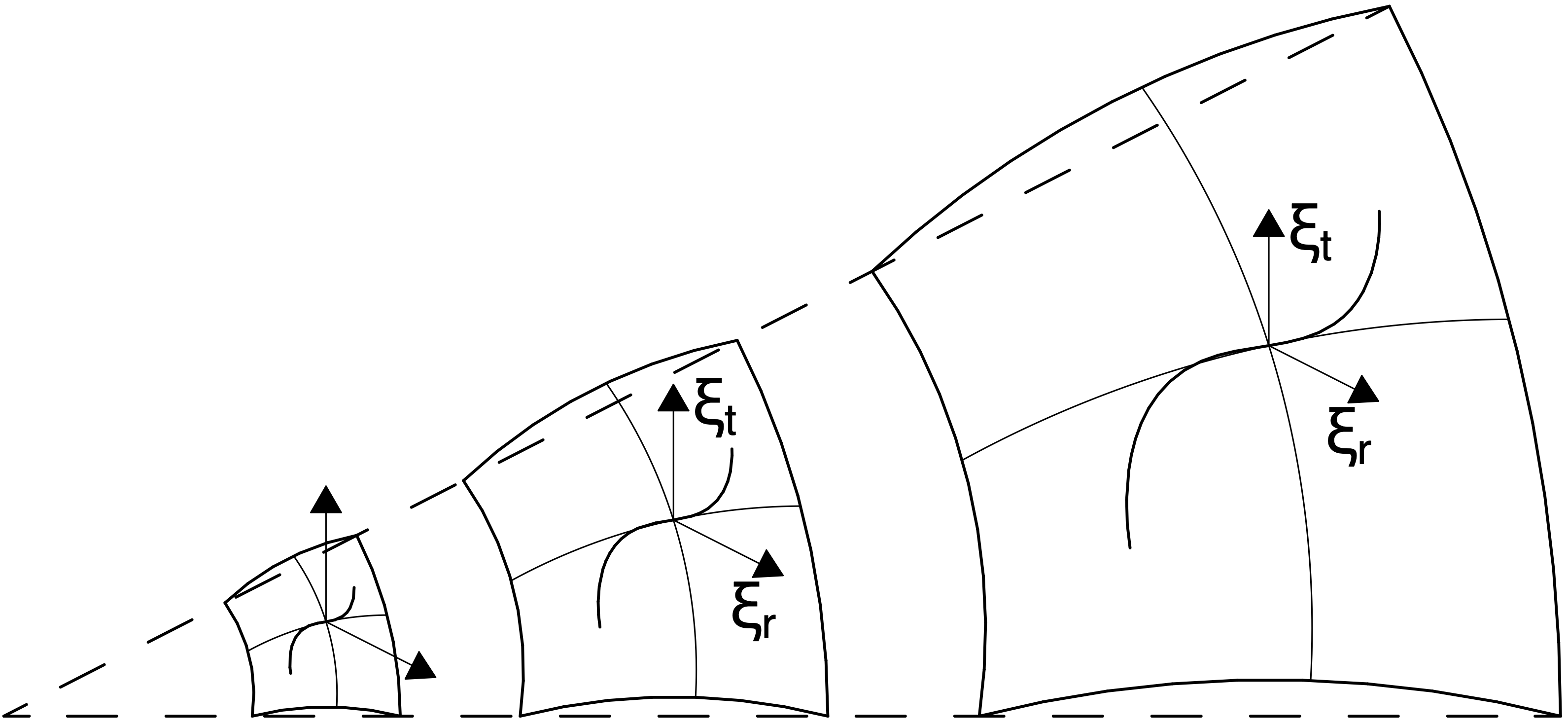}
    \caption{Schematic view of the kinematic self-similar vector fields, with local Killing vector fields, where $\xi_t$ and $\xi_r$ represent the local basis.}
    \label{fig:1}
\end{figure}

Homothetic solutions are a powerful tool for studying the solution of the Einstein Field Equations. Still, the requirement of such symmetry severely restricts the class of equations of state that allow for consistent solutions. Kinematic self-similar solutions have been introduced and used in the context of relativistic hydrodynamics, and their generalization into general relativity is an analogue of the incomplete self-similarity used in Newtonian gravitational theory~\cite{Henriksen1989, Carter1991-tw}. There exists a natural generalization (proposed by Carter and Henriksen in 1989 and 1991) of homothety called kinematic self-similarity, which is defined by the existence of a kinematic self-similar vector field (KSSVF), as seen in  Figure~\ref{fig:1}.
These kinds of vector fields must satisfy the following equations
\begin{align} 
    \mathcal{L}_{\boldsymbol{\xi}} h_{\mu \nu} = 2 \delta h_{\mu \nu}, \label{eq::proj}\\
    \mathcal{L}_{\boldsymbol{\xi}} u_{\mu} = \alpha u_{\mu}, \label{eq::velocitylie}
\end{align}
where $h_{\mu \nu} := g_{\mu \nu}+ u_{\mu} u_{\nu}$ acts as a tangential projector onto $\Sigma_t$ 3-dimensional spacelike hypersurface $(t(x_{\mu}) = \text{const.})$. The $\alpha$ and $\delta$ are dimensionless constants called similarity indices. These indices characterize self-similar transformations. The parameter $\alpha$ represents the relative proportionality factor that governs the dilation rates of the spatial length scale and the increase in the time scale. Taking into account the possible cases, the solutions are:
\newpage
\begin{description}
    \item[0\textsuperscript{th} kind] If both similarity indices are equal to zero, the solution is of the zeroth kind (Killing vectors). 
    \item[1\textsuperscript{st} kind] If the $\alpha/\delta$  ratio is equal to unity, it is named the first kind of self-similarity~\cite{HidekiHarada}.
    \item[2\textsuperscript{nd} kind] If $\delta \neq 0$ and $\alpha \neq 0$, then it is called second-order self-similarity.
    \item[$\infty$\textsuperscript{th} kind]  Lastly, the case where $\delta = 0, \alpha \neq 0$ is an infinite kind of solution. 
\end{description}
Several authors have explored kinematic self-similar perfect-fluid solutions~\cite{PhysRevD.42.1068,Benoit1998,Carr1999}. We aim to investigate how these solutions are modified when viscous terms are taken into account.

\subsection{Viscous Model}

In our analysis, we assumed that the space-time is spherically symmetric, but non-static and then the line element can be written in the following form
\begin{equation} \label{eq::line_element}
\begin{split}
\textrm{d}s^2 = - \textrm{e}^{2\Phi} \textrm{d}t^2 + \textrm{e}^{2\Psi}\textrm{d}r^2 + R^2 \left[\textrm{d}\theta^2 + \Sigma(k,\theta)^2 \textrm{d}\phi^2 \right],
\end{split}
\end{equation}
where, 
\begin{equation} \label{eq:topology}
    \Sigma(k,\theta) =
\begin{cases}
\sin(\theta), \quad  \text{ if} \quad k = 1 ;\\
\theta, \quad \quad  \quad \text{ if} \quad k = 0 ;\\
\sinh(\theta), \quad \text{if} \quad k = -1, 
\end{cases}
\end{equation}
and $\Phi = \Phi(t,r)$, $\Psi=\Psi(t,r)$  and $R =R(t,r)$ are arbitrary functions of time and radial coordinates. We use the exponential form of these functions to ensure positivity and avoid singularities. Furthermore, the exponential form often allows a more direct interpretation of these physical quantities. The function $\Phi(t,r)$ determines the time dilation effects and the gravitational potential and $\Psi(t,r)$ accounts for the spatial curvature in the radial direction. The exponential form of the $R(t,r)$ function is deliberately avoided to maintain its direct physical interpretability and to facilitate a more meaningful comparison with the Friedmann\,--\,Robertson\,--\,Walker\,--\,Lemaître (FRWL) model~\cite{Walker1937, Friedmann1922}. The energy-momentum tensor describes the viscous model in Landau\,--\,Lifschitz\,--\,Eckart flow~\cite{landau1987fluid, PhysRev.58.919}, 
\begin{equation} \label{eq:ViscousEnergyTensor}
  T_{\mu \nu} = (\rho + p + \Pi) u_{\mu} u_{\nu} + (p + \Pi) g_{\mu \nu} + 2 \eta \sigma_{\mu \nu}.
\end{equation}
Introducing $P$, which can be considered as effective pressure modified by the bulk viscosity ($\Pi$), 
\begin{equation} \label{eq::effective_pressure}
 P := p + \Pi =  p - \zeta(\rho, \theta)\theta,   
\end{equation}
where $\zeta(\rho, \theta)$ is the bulk viscosity coefficient and $\theta = u^{\mu}_{;\mu}$ is the expansion scalar. The shear viscosity parameter $\eta$, quantifies the dissipative stress produced by anisotropic or shape-changing deformations of the fluid that occur at constant volume. We have used the linear barotropic equation of state, $p=w_0\rho$, which is compatible with the kinematic self-similarity conditions~\cite{PhysRevD.66.027501}. We assume that the $w_0$ parameter is $-1 \leq w_0 \leq 1$.  One can define the $w(t,r)$ effective Equation of State (EoS) parameter between the density and the effective pressure $P=w(t,r)\rho$ as 
\begin{equation} \label{eq::w_definition}
    w(t,r) = w_0 - \dfrac{\zeta(\rho, \theta)}{\rho}\textrm{e}^{-\Phi}\left(\dot{\Psi}+2\dfrac{\dot{R}}{R} \right).
\end{equation}
The effective pressure $P(\xi)$ must satisfy $\LieXi P = - 2 \alpha P$ based on the argument made in Appendix~\ref{sec::AppendixLie}. It can be seen from the definition of Eq.~\eqref{eq::effective_pressure} that both terms in the sum are required to satisfy the Lie condition, thus $\LieXi \Pi = - 2 \alpha \Pi$. Since $\theta$ can be expressed as the covariant derivative of $u_{\mu}$ and $\LieXi u_{\mu} = \alpha u_{\mu}$, the $\zeta(\rho, \theta)$ should scale as  $\LieXi \zeta( \rho, \theta) = \alpha \zeta(\rho, \theta)$. Solutions both exhibit proper scaling and covariant nature and depend only on one of the dynamical variable are $\zeta(\rho) \propto \rho^{1/2}$ and $\zeta(\theta) \propto \theta$. Choosing $\zeta(\rho) \propto \rho^{1/2}$ has been widely employed both in cosmology, where it enables analytic treatments of anisotropic viscous models, and in neutron-star physics, where similar fractional density dependences naturally arise in weak-interaction-mediated damping of stellar oscillations~\cite{Acquaviva_2015, Carlevaro}. The second approach is to let $\zeta$ depend on the expansion scalar $\theta$, which highlights the intrinsic thermodynamic role of viscosity: the co-moving expansion of the Universe itself generates bulk viscous effects~\cite{ARORA20221, PhysRevD.111.083540}, then the shear viscosity tensor goes as, 
\begin{equation}
       \sigma_{\mu \nu} = \dfrac{1}{2} (h_{\mu \alpha} \nabla^{\alpha} u_{\nu} + h_{\nu \alpha} \nabla^{\alpha} u_{\mu}) - \dfrac{1}{3} h_{\mu \nu} \theta \ .
\end{equation}
The decomposition of the 4-velocity and its covariant derivative can be found in Appendix~\ref{Sec::Append2}, and the Lie derivatives of the stress-energy-momentum tensor can be found in Appendix~\ref{sec::AppendixLie}. We adopt a co-moving frame, where the 4-velocity takes the form of
\begin{equation}
    u_{\mu} = (\textrm{e}^{-\Phi},0,0,0).
\end{equation}
Thus, one can obtain the following expression for the expansion scalar, 
\begin{equation}
    \theta = \textrm{e}^{-\Phi} \left( \Dot{\Psi} + 2 \dfrac{\Dot{R}}{R} \right),
\end{equation}
and the non-zero components of the shear viscosity tensor
\begin{align}
    \sigma^r_{r} & = \textrm{e}^{- \Phi} \dfrac{2}{3} \left( \Dot{\Psi} - \dfrac{\Dot{R}}{R} \right), \\
    \sigma_{\phi}^{\phi} & = \sigma^{\theta}_{\theta} = - \textrm{e}^{-\Phi} \dfrac{1}{3} \left( \Dot{\Psi} -  \dfrac{\Dot{R}}{R} \right).
\end{align}
Utilizing the derived expressions, the shear viscosity scalar can be defined as follows
\begin{equation}
    \sigma^2 = \dfrac{1}{2}\sigma_{\mu \nu}\sigma^{\mu \nu} = \textrm{e}^{-2\Phi} \dfrac{1}{3}\left( \Dot{\Psi} - \dfrac{\Dot{R}}{R} \right)^2.
\end{equation}
Consequently, the Einstein field equation expressed in terms of Misner\,--\,Sharp\,--\,Hernandez (MSH) mass and the Bianchi identity have the form,
\begin{align}
    -P' &=  (\rho + P) \Phi' - (2\eta\sigma^{r}_r)'  \notag \\ 
    &- 6 \eta \sigma_r^r \left(\dfrac{R'}{R} - \dfrac{\Phi'}{3}\right), \label{eq::Cons1}\\
    -\Dot{\rho} \textrm{e}^{-\Phi} &= (\rho + P) \theta - 4 \eta \sigma^2,  \label{eq::Cons2}\\
     m' &= 4\pi \rho R'R^2, \label{eq::EFE1} \\
     \Dot{m} & = - 4 \pi (P - 2 \eta \sigma^r_r) \Dot{R}R^2,  \label{eq::EFE2}
\end{align}
\begin{equation}
    \Dot{R}' = \Phi'\Dot{R} + \Dot{\Psi}R',  \label{eq::EFE3}
\end{equation}
where the $m = m(t,r)$ is the MSH mass defined via the equations of \cite{Misner1964,Hernandez1966},
\begin{equation}
    g_{rr} = \left[ 1 + \left(\textrm{e}^{-\Phi}\Dot{R}\right)^2 - \dfrac{2m}{R} \right]^{-1}  R'^2.
\end{equation}
The relevant components of the Ricci tensor and Ricci scalar are described in Appendix~\ref{sec::appendB}. Also, the remaining (mixed EFEs) Eqs.~\eqref{eq::aux_0}-\eqref{eq::aux_1} act as supplementary equations for the analysis
\begin{widetext}
\begin{align}
    \kappa \rho & =\textrm{e}^{-2\Phi}R^{-2} \left[ \dot{R} \left( \dot{R} + 2 R \dot{\Psi} \right) \right]  + R^{-2}  + \textrm{e}^{-2\Psi}R^{-2}  \left( -2 R R' \Psi' + R'^2 + 2 R R''  \right), \label{eq::aux_0} \\
   \kappa (P -2 \eta \sigma_r^r) & =  \textrm{e}^{-2\Psi} R^{-2} \bigg[ 2 R R' \Phi' + R'^2 -\textrm{e}^{2(\Psi - \Phi)}  \left( \textrm{e}^{2\Phi} - 2 R \dot{R} \dot{\Phi} + \dot{R}^2 + 2 R \ddot{R} \right) \bigg], \\
    \kappa (P -2 \eta \sigma_{\theta}^{\theta}) & = \textrm{e}^{-2 (\Psi + \Phi)} R^{-1} \bigg[ \textrm{e}^{2\Phi} \bigg( \big( R' + R \Phi' \big) \big( \Phi' - \Psi' \big) + R'' + R \Phi''\bigg) + \textrm{e}^{2\Psi} \bigg( \big( \dot{R} + R \dot{\Psi} \big) \big( \dot{\Phi} - \dot{\Psi} \big) - \ddot{R} - R \ddot{\Psi} \bigg) \bigg]. \label{eq::aux_1}
\end{align}
\end{widetext}
One can assume that the particle number is conserved, which is defined 
\begin{equation} \label{eq::particle_conserved}
    \nabla_{\mu} N^{\mu} = \nabla_{\mu} (nu^{\mu}) = 0 \Rightarrow \textrm{e}^{-\Phi}\Dot{n} + n \theta = 0,
\end{equation}
where $n=n(t,r)$ and $N^{\mu}=N^{\mu}(t,r)$ are the particle number density and particle current respectively. As a consequence of the definition of $h_{\mu \nu}$ projection tensor and the condition in Eq.~\eqref{eq::proj}, spatial lengths scale with weight $\delta$, since the particle number density, defined as the number per spatial volume unit, must scale with $\delta$. Hence, 
\begin{equation}
    \LieXi N_{\mu} = (\LieXi n) u_\mu + n \LieXi u_{\mu} = (\alpha - \delta)N_{\mu}, 
\end{equation}
and therefore $N^{\mu}$ satisfies the homothety condition.

\section{Results}
\label{sec:results}

Spherical-symmetric, time-dependent scenarios for bulk viscosity are presented in this section for various cases. Foremost, we incorporate bulk viscosity, examining its impact on dynamic evolution and its asymptotic behavior. It is known that, for a general spherically symmetric space-time, the kinematic self-similar vector field is given by the following formula 
\begin{equation} \label{eq::SphKSS}
    \boldsymbol{\xi}(t,r) = h_1(t,r) \dfrac{\partial}{\partial t} + h_2(t,r) \dfrac{\partial}{\partial r},
\end{equation}
where $h_1(t,r)$ and $h_2(t,r)$ are functions of time and spatial coordinate. One solution is called {\sl orthogonal} to the fluid flow if  $h_1(t,r) = 0$, and it is {\sl parallel} if $h_2(t,r) = 0$; otherwise, it is called {\sl tilted}. 
\subsection{Tilted Case}
We begin by considering the most general scenario, where the KSSVFs are tilted and therefore not aligned either parallel or orthogonal. If we exclude the infinite kind of self-similarity, then the $\delta$ similarity parameter can be set to unity. Then the $h_1(r,t)$ and $h_2(r,t)$ functions can be written in terms of $\alpha$ and $\beta$ constants~\cite{brandt}
\begin{equation} \label{eq::kinematic_vector_field}
       \boldsymbol{\xi}(r,t) = (\alpha t+\beta) \dfrac{\partial}{\partial t} + r \dfrac{\partial}{\partial r}.
\end{equation}
Hence, if the $\alpha$ is unity, one can set the $\beta$ to zero, and the kinematic self-similar vector field is equivalent to the first kind of self-similarity. In the case of the type of zeroth self-similarity, the $\alpha$ vanishes, and then the $\beta$ is set to unity. For $\alpha$ neither zero nor unity, the $\beta$ again can be fixed to zero, and the self-similarity becomes the second kind. It is summarized in Table~\ref{tab:self_similarity}.
\begin{table}[H]
    \centering
    \begin{tabular}{ccc}
        \hline
     
           \textbf{Self-similarity} & \textbf{Similarity} & \textbf{Similarity} \\
              \textbf{type} & \textbf{variable} & \textbf{constants} \\
        \hline
        \hline
        \(0^{\text{th}}\) kind  & \(\xi = r \mathrm{e}^{-t}\) & $\alpha = 0$, $\beta=1$ \\
        \(1^{\text{st}}\) kind  & \(\xi = r/t\) & $\alpha=1$, $\beta=0$ \\
        \(2^{\text{nd}}\) kind  & \(\xi = r(\alpha t)^{-1/\alpha}\) & $\alpha\neq0,1$, $\beta=1$\\
        \( \infty^{\text{th}}\) kind  & \(\xi = r/t\) & \,--\, \\
        \hline
    \end{tabular}
    \caption{Classifications of self-similarity types and their corresponding similarity variable        ~\cite{HidekiHarada}.}
    \label{tab:self_similarity}
\end{table}
The unknown metric functions we are looking for can be expressed in terms of the similarity variable based on dimensional analysis
\begin{align} 
    R(t,r) = r \mathcal{R}(\xi), \quad \Phi(t,r) = \Phi(\xi), \quad
    \Psi(t,r) = \Psi(\xi).\label{eq::canonical_similar_form}
\end{align}
and consequently the metric, 
\begin{equation} \label{eq::line_element_2nd}
    \mathrm{d}s^2 = -\textrm{e}^{2 \Phi}\mathrm{d}t^2 + \textrm{e}^{2 \Psi}\mathrm{d}r^2 + r^2 \mathcal{R}^2\mathrm{d} \Omega^2,
\end{equation}
If the $\alpha \neq 0$, the particle number density takes the canonical self-similar form of,
\begin{equation} \label{eq::particle_density_self_similar}
    n(t,r) = t^{-\delta/\alpha} \Tilde{N}(\xi).
\end{equation}
Also if the $\alpha =0$ and the self-similarity variable became $\xi = r\mathrm{e}^{-t}$, the previous expression varies as
\begin{equation} \label{eq::part_num_zeroth}
    n(t,r) = \mathrm{e}^{-\delta t} \tilde{N}(\xi).
\end{equation}
If the $\delta = 0$ and $\alpha = 1$, the kinematic self-similar vector becomes an infinite kind, and the vector field is
\begin{equation} \label{eq::inifity_tilted}
    \boldsymbol{\xi}(r,t) = t \dfrac{\partial}{\partial t} + r \dfrac{\partial}{\partial r},  
\end{equation}
with self-similarity variable of $\xi = r/t$ and the metric functions are
\begin{equation}
    R(t,r) = \mathcal{R}(\xi), \quad \Phi(t,r) = \Phi(\xi), \quad \textrm{e}^{\Psi(t,r)} = \textrm{e}^{\Psi(\xi)}/r,
\end{equation}
which are obtained by solving equations of Eqs.~\eqref{eq::proj}-\eqref{eq::velocitylie} for a general spherically symmetric space-time, if the KSSVF is Eq.~\eqref{eq::inifity_tilted} analyzed by Sintes {\it et al}  \cite{Sintes2001-fy}. For the infinite kind of self-similarity, the expression for particle number density in Eq.~\eqref{eq::particle_density_self_similar} simplifies as $n(t,r) = \Tilde{N}(\xi)$.
\subsubsection{Self-similarity of Second Kind}
It follows from Einstein's Field Equation that a set of ordinary differential equations is constructed for the energy density, the effective pressure, and the Misner\,--\,Shapiro mass
\begin{align} 
    \kappa \rho & = \dfrac{1}{r^2} \left[ \rho_0(\xi) + \dfrac{r^2}{t^2} \rho_2(\xi)\right],  \label{eq::EquationOrder1}\\
    \kappa P & = \dfrac{1}{r^2} \left[ P_0(\xi) + \dfrac{r^2}{t^2} P_2(\xi)\right],\label{eq::EquationOrder2}\\ 
    2m & = r \left[ m_0(\xi) + \dfrac{r^2}{t^2} m_2(\xi)\right]. \label{eq::EquationOrder3}
\end{align}
This separation is a direct consequence of the kinematic self-similar symmetry, which permits the replacement of spatial and temporal derivatives by derivatives with respect to the similarity variable, $\xi = r(\alpha t)^{-1/\alpha}$. As a result, the EFEs and matter field equations naturally split into components proportional to $\mathcal{O}[(r/t)^0]$ and $\mathcal{O}[(r/t)^2]$,  which must be satisfied independently \cite{Cahill1971}. The EFEs and the conservation equations in the Eq.~\eqref{eq::Cons1}-\eqref{eq::EFE3} transforms accordingly, 
\begin{align}
    (\rho_0 + P_0) \Phi' & = 2P_0 - P_0', \label{eq::2ndEq1}\\
    (\rho_2 + P_2) \Phi' & = - P_2', \label{eq::2ndEq2}
\end{align}
\begin{align}
    - \rho_0' \mathcal{R} & = (\rho_0 + P_0) (\Psi' \mathcal{R} + 2 \mathcal{R}'),\label{eq::2ndEq3}  \\
    -(2 \alpha \rho_2 + \rho_2') \mathcal{R} & =  (\rho_2 + P_2) (\Psi' \mathcal{R} + 2 \mathcal{R}'), \label{eq::2ndEq4}
\end{align}
\begin{align}
    m_0 + m_0' & =  \rho_0 \mathcal{R}^2 (\mathcal{R}+ \mathcal{R}'), \label{eq::2ndm1}\\
    3 m_2 + m_2' & = \rho_2 \mathcal{R}^2 (\mathcal{R} + \mathcal{R}'), \label{eq::2ndm2}
\end{align}
\begin{align}
    m_0' = - P_0 \mathcal{R}^2 \mathcal{R}', \label{eq::2ndconds_1}\\
    2 \alpha m_2 + m_2' = -P_2 \mathcal{R}^2 \mathcal{R}',\label{eq::2ndcons_2}
\end{align}
\begin{equation}
    \mathcal{R}''+ \mathcal{R}' = \mathcal{R}' \Phi' + (\mathcal{R} + \mathcal{R}')\Psi', \label{eq::2ndrt}
\end{equation}
and for the Misner\,--\,Sharp\,--\,Hernandez mass,
\begin{align}
    m_0 &= \mathcal{R} [1 - \textrm{e}^{-2\Psi} (\mathcal{R} + \mathcal{R}')^2 ], \label{eq::misner_2nd_1}\\
    \alpha^2 m_2 & = \mathcal{R}' \mathcal{R}^{2} \textrm{e}^{-2\Phi}. \label{eq::misner_2nd_2}
\end{align}
The auxiliary equations are, 
\begin{align}
    \cR'^2 + 2 \Psi' \cR \cR' & = \alpha^2 \rho_2 \cR^2 e^{2\Phi},\label{eq::aux_2nd_0} \\
     \cR (2 \cR'' + 4 \cR') & = 2 \Psi' (\cR^2 + \cR \cR') - \notag \\
     - (\cR'^2 + \cR^2) & + \mathrm{e}^{2 \Psi} (1 - \rho_0 \cR^2),  \label{eq::aux_2nd_1}\\ 
    -2 \cR ( \cR'' + (\alpha - \Phi') \cR') & = \alpha^2 P_2 \cR^2 e^{2\Phi} + \cR'^2 ,\label{eq::aux_2nd_3}\\
    (\cR' + \cR)(\cR' + \cR + 2 \Phi' \cR) & = \textrm{e}^{2\Psi}(1 + P_0\cR^2), \label{eq::aux_2nd_2}
\end{align}
where the $(')$ denotes the derivative concerning the logarithmic of the similarity variable $\ln \xi$. Similarly, an equation for the particle number density can be obtained if one substitutes the expression from Eqs.~\eqref{eq::canonical_similar_form} and~\eqref{eq::particle_density_self_similar} into Eq.~\eqref{eq::particle_conserved}, which will result
\begin{equation} \label{eq::particle_number_density}
\dfrac{\Tilde{N}'}{\Tilde{N}} = - \dfrac{1}{\alpha}\left[ 2\dfrac{\cR'}{\cR}+ \Psi' + \delta\right].
\end{equation}
Integration with respect to $\ln \xi$, will lead to
\begin{equation}
    \Tilde{N} = C_{N} \xi^{-\delta/\alpha} \cR^{-2/\alpha} \mathrm{e}^{-\Psi/\alpha}.
\end{equation}
The detailed derivation of the obtained solution can be found in Appendix~\ref{sec::appendixE}. In the first scenario, where $\rho_0$ vanishes, we proved that the $\zeta(\rho) \propto \rho^{1/2}$ bulk viscosity condition is not compatible with the self-similarity condition. However, in the case of $\zeta(\theta) \propto \theta$, the non-constant metric functions in Eq.~\eqref{eq::line_element_2nd} are
\begin{align}
    \mathcal{R}(\xi) &= \mathcal{R}_0 \left[ 3(1-\tilde{U}_0)\,\xi^{C_U} + (1-\tilde{U}_0)^{C_U} \right]^{1/3}\\
    \Psi(\xi) & = \ln \Bigg( \cR_0 \xi ^{C_U/3} \bigg( 3-2\alpha+(1-\tilde{U}_0)^{C_U}\xi^{-C_U} \bigg) \notag \\
    &\times \left( 3(1-\tilde{U_0}) + (1-\tilde{U_0})^{C_U}\xi^{-C_U}\right)^{-2/3}\Bigg),
\end{align}
This solution aligns with the self-similar dust models classified by Carr and Coley \cite{PhysRevD.62.044022}, particularly those belonging to the particular version of Kantowski\,--\,Sachs model~\cite{10.1063/1.1704952} with an appropriate set of constants. Since $\xi = r(\alpha t)^{-1/\alpha}$ and $R(t,r) = r\,\mathcal{R}(\xi)$, the obtained solution reduces to $\mathcal{R}(\xi) \propto \xi^{-1}$ when $C_U = -3$. In this case, the area radius behaves as $R(t,r) \propto t^{-C_U/\alpha}$, implying that the two-sphere radius does not depend on the spatial coordinates, as in the Kantowski\,--\,Sachs metric. Under this exact condition, the metric function $\Psi(\xi)$ becomes a function of $t$ only, while $\Phi(\xi)$ remains constant. Notice that this power-law behavior holds only asymptotically, in the long-time limit ($\xi \to 0$), where the power-law term becomes dominant in the metric functions. This case excludes the first-kind kinematic self-similar solution, since $C_U = (3\tilde{U}_0 - 2\alpha)(1 - \tilde{U}_0)$, cannot be satisfied for any finite value of $\tilde{U}_0$. In the next scenario, which is the non-trivial version of the $\rho_2=0$, the line element becomes
\begin{align}
\mathrm{d}s^2 & = - r^{2 c_{\Phi}} (\alpha t)^{-\frac{2 c_{\Phi}}{\alpha}} \mathrm{d}t^2 
+ r^{-\lambda_{0,r}} (\alpha t)^{\frac{\lambda_{0,r}}{\alpha}} \mathrm{d}r^2 
 \\ \notag 
& +\mathcal{R}_0^2 \, r^{2 + 2 \lambda_{0,r}} (\alpha t)^{-\frac{2 \lambda_{0,r}}{\alpha}} \mathrm{d}\Omega^2,
\end{align}
By setting $c_{\Phi} = 0$ which implies, $\alpha = 3/2$ and consequently $\lambda_{0,r} = -1$, the metric reduces to the standard FRWL from. Asymptotically FRWL solutions have been discussed in the context of PBH formation by Carr and Hawking~\cite{Hawking1974}, and subsequently Bicknell and Henriksen ~\cite{1978Bicknell, osti_6856773}. In the case of the so-called trivial solution, the only non-constant term is 
\begin{equation}
    \mathrm{e}^{2\Phi} = \xi^{\dfrac{4w_0}{(1+w_0)}},
\end{equation}
and the metric follows as 
\begin{equation}
    \mathrm{d}s^2  =  -\xi^{\dfrac{4w_0}{(1+w_0)}} \mathrm{d}t^2 + \Psi_0^2 \mathrm{d}r^2  +  \cR_0^2 r^2\mathrm{d}\Omega^2,
\end{equation}
which is called a static singular solution, due to its singularity at $t=0$. 
\subsubsection{Self-similarity of First Kind (Homothetic solution)}
As in the previous case, the EFEs and the conservation equations in Eqs.~\eqref{eq::Cons1}-\eqref{eq::EFE3} can be expressed in the same form as those obtained for self-similarity of the second kind. One can easily notice that this differential equation system is too restrictive. However, the homothetic solution can still be obtained as a limiting case ($\alpha=1)$ of the second-type self-similar solution.
\subsubsection{Self-similarity of Zeroth Kind}

In the case of self-similarity of the zeroth kind, the Einstein equations imply similar constraints as it was derived in Eq.~\eqref{eq::EquationOrder1}-\eqref{eq::EquationOrder3},
\begin{align} 
    \kappa \rho & = \dfrac{1}{r^2} \bigg[ \rho_0(\xi) + r^2 \rho_2(\xi)\bigg],  \label{eq::EquationOrder4}\\
    \kappa p & = \dfrac{1}{r^2} \bigg[ P_0(\xi) + r^2 P_2(\xi)\bigg],\label{eq::EquationOrder5}\\ 
    2m & = r \bigg[ m_0(\xi) + r^2 m_2(\xi)\bigg], \label{eq::EquationOrder6}
\end{align}
and the $\xi=r\mathrm{e}^{-t}$ similarity variable defined in Table.~\ref{tab:self_similarity}. The obtained equations are mostly similar to the homothetic case. In particular, Eqs.~\eqref{eq::2ndEq1}-\eqref{eq::2ndEq2}, \eqref{eq::2ndm1}-\eqref{eq::2ndm2} (with $\alpha=1$), and \eqref{eq::2ndrt}, as well as the equation for the Misner\,--\,Sharp\,--\,Hernandez mass, remain equivalent. The differences arise in the remaining equations, which are as follows:
\begin{align}
     -\rho_0' \mathcal{R} & = (\rho_0 + P_0) (\Psi' \mathcal{R} + 2 \mathcal{R}'),  \label{eq::0th_order_rho0} \\
    -\rho_2' \mathcal{R} & = (\rho_2 + P_2) (\Psi' \mathcal{R} + 2 \mathcal{R}') ,\label{eq::0th_order_rho2}\\
    -m_0' & = P_0 \mathcal{R}^2 \mathcal{R}', \label{eq::0th_m0_def}\\ 
    -m_2' & = P_2 \mathcal{R}^2 \mathcal{R}'.  \label{eq::0th_m2_def}
\end{align}
The auxiliary equations of Eqs.~\eqref{eq::aux_2nd_0}-\eqref{eq::aux_2nd_2} and particle number density equations of Eq.~\eqref{eq::particle_density_self_similar} are reduced to the form, where $\alpha$ becomes unity. Thus, we have received the solution
\begin{equation} \label{eq::part_num_sol}
    \Tilde{N}(\xi) = C_{\tilde{N}} \xi^{-\delta} \cR^{-2}\mathrm{e}^{-\Psi}.
\end{equation}
For the $\rho_0=0$ case, we obtained the following metric,
\begin{equation}
    \mathrm{d}s^2 = -\Phi_0^2\mathrm{d}t^2 + 4 \cR_0^2 \mathrm{e}^{-2t} \mathrm{d}r^2 + \cR_0^2r^2\mathrm{e}^{-2t}\mathrm{d}\Omega^2.
\end{equation}
The detailed derivation can be seen in Appendix~\ref{sec::sec::zeroth}. However, the solutions reduce to the perfect fluid solution, since the $\theta$ and $\rho_2$ become constant. If $\rho_2$ vanishes, we will arrive at the following metric,
\begin{equation}
    \mathrm{d}s^2 = \mathrm{e}^{2c_{\Phi}}(r\mathrm{e}^{-t})^{2w_0/(1+w_0)} \mathrm{d}t^2 + \dfrac{c_{\cR}^2}{c_{\cR}-c_{m}} \mathrm{d} r^2 + r^2 c_{\cR}^2 \mathrm{d}\Omega^2.
\end{equation}
If someone applies the appropriate coordinate transformation of $\tilde{r} = c_{\cR}r$ and
\begin{equation}
      \tilde{t} = -\left(\dfrac{1+w_0}{w_0}\right)\mathrm{e}^{c_{\Phi}}c_{\cR}^{-w_0/(1+w_0)}\mathrm{e}^{2w_0/(1+w_0)t}.
\end{equation}
Under this transformation rule, the metric becomes, 
\begin{equation}
    \mathrm{d}s^2= \tilde{r}^{2w_0/(1+w_0)} \mathrm{d}t^2 + \dfrac{1}{c_{\cR}-c_{m}}\mathrm{d}\tilde{r}^2+\tilde{r}^2 \mathrm{d}\Omega^2,
\end{equation}
which is a similar static singular solution to the obtained solution for the self-similarity of the second-kind case. Analogous to the previous case, the detailed analysis in the Appendix shows that the $w(\xi)$ must remain constant, and the viscosity term should disappear.
\subsubsection{Self-similarity of Infinite Kind}
Under the assumption of infinite kind of self-similarity, the Einstein field equations split into terms of order 
$(1/t)^2$ and $(1/t)^0$. The energy density, the effective pressure and the MSH mass are decomposed accordingly
\begin{align} 
    \kappa \rho & =  \bigg[ \rho_0(\xi) + \dfrac{1}{t^2} \rho_2(\xi) \bigg] , \label{eq::infti_dens} \\ 
    \kappa p & =  \bigg[ P_0(\xi) + \dfrac{1}{t^2} P_2(\xi) \bigg],  \label{eq::infti_press} \\ 
    2m & = \bigg[ m_0(\xi) + \dfrac{1}{t^2} m_2(\xi) \bigg],  \label{eq::infti_mass}
\end{align}
where the self-similarity variable $\xi=r/t$. Consequently, the definition of the MSH mass takes the form of 
\begin{align}
    m_0 & = \cR \left(1 - \mathrm{e}^{-2\Psi}\cR'^2 \right), \\
    m_2 & = \cR \cR'^2 \mathrm{e}^{-2\Phi}. 
\end{align}
Such that spatial mass conservation equations remain similar to the Eqs~\eqref{eq::2ndconds_1}-\eqref{eq::2ndcons_2} (with $\alpha=1$) and the temporal part becomes
\begin{align}
    m_0' & = \rho_0 \cR^2 \cR', \label{eq::infinity_mass_cons_0} \\
    m_2' & = \rho_2 \cR^2 \cR'. 
\end{align}
The energy conservation $T_{\mu \nu;\nu} = 0$ becomes 
\begin{align}
    (\rho_0 + P_0) \Phi' & = - P_0', \label{eq::infinity_rho_0}\\
    (\rho_2 + P_2) \Phi' & = - P_2', \label{eq::infinity_rho_2}\\
    \rho_0'\cR & =  -(\rho_0 + P_0) (\Psi' \mathcal{R} + 2 \mathcal{R}'), \\
    (2\rho_2 + \rho_2') \mathcal{R} & = -(\rho_2 + P_2) (\Psi' \mathcal{R} + 2 \mathcal{R}'), \label{eq::infinity_eq_4}
\end{align}
The $(r,t)$ component of the Einstein Field Equations takes the simple form of,
\begin{equation} \label{eq::rt_equation_infinite}
    \cR'' = \cR'(\Phi' + \Psi'), 
\end{equation}
and the auxiliary equations are
\begin{align}
    \rho_2 \cR^2 \mathrm{e}^{2\Phi} & = \cR' (2 \Psi' \cR + \cR'), \label{eq::infinity_aux_1}\\
    2 \cR [\cR'' + (1 - \Phi') \cR'] & = - (\cR'^2 + P_2 \cR^2\mathrm{e}^{2\Phi}), \label{eq::infity_aux_2} \\
    (1-\rho_0\cR^2) \mathrm{e}^{2\Psi}  & = 2\cR(\cR''-\Psi'\cR') + \cR'^2, \label{eq::infinity_aux_3}\\
    (1 + P_0 \cR^2) \mathrm{e}^{2\Psi} & = \cR' (\cR' + 2 \Phi' \cR). \label{eq::infinity_aux_4}
\end{align}
The obtained solution for $\Tilde{N}(\xi)$ in Eq.~\eqref{eq::part_num_sol} is still holds for this self-similarity case. We have verified in Appendix~\ref{sec::sec::infinite_tilted} subsection that no viscous solution exists if $\rho_0\neq0$, since it reduces to the perfect fluid solution. Also, we demonstrated that $\rho_0 = 0$ implies, the given differential equation system does not admit a general analytic solution, but its fixed points can still be determined explicitly. The linear stability analyses resulted in the system having one singular and one stable fixed point. Around the stable fixed point, the relevant EFEs lead to the following metric form,
\begin{equation}
    \mathrm{d}s^2 = - \Phi_0^2 \mathrm{d}t^2 + U^2_2 \cR_0^2 \left( \dfrac{r}{t}\right)^{2U_2} \mathrm{d}r^2 + \cR^2_0 \left( \dfrac{r}{t} \right)^{2U_2} \mathrm{d} \Omega^2,
\end{equation}
where $U_2=-1/8(4+ 3 c_{P_2}\Phi_0^2)$ is a constant. One can make an appropriate variable transformation as $t \rightarrow t' = \Phi_0t$ and $r \rightarrow r'= r^{U_2+1}(U_2+1)^{-1}$ and the metric becomes
\begin{equation}
    \mathrm{d}s^2 = -\mathrm{d}t'^2 + \cR_0^2 \left( \dfrac{\Phi_0}{t}\right)^{2U_2}  \left( U_2^2 \mathrm{d}r'^2 + r'^{2U_2} \mathrm{d} \Omega^2\right),  
\end{equation}
which is equivalent with the conformally flat FRWL metric if $U_2=1$ and $\Sigma(\theta,k)=\theta$.
\subsection{Parallel Case}
In the case when the kinematic vector field is parallel to the fluid flow, the Eq.~\eqref{eq::kinematic_vector_field} becomes
\begin{equation} \label{eq::parallel_kssvfs}
    \boldsymbol{\xi}(t,r)  = t \dfrac{\partial}{\partial t},
\end{equation}
and the self-similar variable $\xi$ becomes $r$. Similarly, we can define the line element in the co-moving frame 
\begin{equation}
    \mathrm{d}s^2 = - t^{2(\alpha-1)}\mathrm{e}^{2\Phi(\xi)}\mathrm{d}t^2+ t^2 \mathrm{d}r^2 + t^2\cR^2(\xi) \mathrm{d}\Omega^2, 
\end{equation}
where $\alpha$ self-similarity index $\alpha \neq 0$ and $\alpha \neq 1$ for self-similarity of second kind, and $\alpha=0$ for self-similarity of zeroth kind. The line element will be modified for the infinite kind, such as
\begin{equation} \label{eq::parallel_metric_infinity}
    \mathrm{d}s^2=-\mathrm{e}^{2\Phi(\xi)} \mathrm{d}t^2 + \mathrm{d}r^2 + \cR^2(\xi) \mathrm{d}\Omega^2. 
\end{equation}
\subsubsection{Self-similarity of the First, Second, and Zeroth Kind}
The form of Einstein's field equations, as well as in the tilted solutions, implies the form of density, effective pressure, and MSH-mass, which will result in 
\begin{align}
    \kappa \rho & = t^{-2}\rho_0(\xi)+t^{-2\alpha} \rho_2(\xi),\\
    \kappa P & = t^{-2}P_0(\xi)+t^{-2\alpha} P_2(\xi),\\ 
    2m & = tm_0(\xi)+t^{3-2\alpha}m_2(\xi).
\end{align}
Since the EFEs are independently satisfied at both orders $\mathcal{O}(t^0)$ and $\mathcal{O}(t^{2-2\alpha})$. The relevant equations will be
\begin{align}
    (\rho_0+P_0)\Phi' & = - P'_0  \quad \mathrm{and}, \quad (\rho_2+P_2)\Phi' = - P'_2,\\
    - \rho_0 & = 3 P_0, \quad  \mathrm{and} \quad (2\alpha-3)\rho_2 = 3P_2. \label{eq::parallel_conservation}
\end{align}
The Misner\,--\,Sharp\,--\,Hernandez masses take the form of
\begin{equation}
    m_0 = \cR (1 -\cR'^2), \quad \and \quad m_2 = \cR^3 e^{-2\Phi},
\end{equation}
and the time and spatial conservation equations can be expressed as
\begin{align}
    m_0 = -P_0 \cR^3, \quad &\textrm{and} \quad (3-2\alpha)m_2 = -P_2\cR^3,\\ \quad m_0' = \rho_0\cR^2\cR' \quad &\textrm{and}, \quad m_2' = \rho_2\cR^2\cR'.
\end{align}
The Eq.~\eqref{eq::EFE3} and the auxiliary equations will take the form of
\begin{align}
    \cR\Phi'& = 0, \label{eq::parallel_eq_rt}\\
    2 \cR'' \cR + \cR'^2 & = 1 - \rho_0 \cR^2,\\
    \cR'^2 + 2 \Phi' \cR' \cR & = 1 + P_0\cR^2 ,\\
    \rho_2 & = 3e^{-2\Phi} ,\\
    P_2 & = (2\alpha-3)e^{-2\Phi},
\end{align}
where the $(')$ denotes derivation with respect to $r$, since $\xi=r$. It can be trivially seen from Eqs.~\eqref{eq::parallel_conservation} that $w(\xi)$ must be constant, and thus, viscous solutions cannot exist in the case of parallel zeroth-, first-, and second-kind of self-similarity. Hence, in the asymptotic regime, such systems can be effectively described as perfect fluids. 

\subsubsection{Self-similarity of Infinite Kind}

If the $\delta=0$, the kinematic self-similar vector field remains as it was defined in Eq.~\eqref{eq::parallel_kssvfs}, as well as the invariant line element in Eq.~\eqref{eq::parallel_metric_infinity}. Consequently, the form of the Einstein equation will constrain the form of the matter fields, which are 
\begin{equation}
    2m = \Tilde{m}(\xi), \quad \kappa\rho = \Tilde{\rho}(\xi) \quad \mathrm{and} \quad \kappa P = \Tilde{P}(\xi).
\end{equation}
It implies that the Einstein Field Equations (Eqs.~\eqref{eq::Cons1}-\eqref{eq::aux_1}) will be
\begin{align}
    (\Tilde{\rho} + \Tilde{P}) \Phi' & = - P',\\
    \Tilde{m} & = \cR (1-\cR'^2),\\
    \Tilde{m}'& = \Tilde{\rho}\cR'\cR^2,\\
    2\cR''\cR + \cR'^2 & = 1 - \Tilde{\rho}\cR^2,\\
    \cR'(2\Phi'\cR+\cR') &= 1 +\Tilde{P}\cR^2.
\end{align}
One can see that if $w(\xi)$ does not depend on time, then the viscous term in Eq.~\eqref{eq::w_definition} should vanish. As a result, the solution will reduce to the exact dimensionless form of the static Tollman\,--\,Oppenheimer\,--\,Volkoff-equation for perfect fluid \cite{DELGATY1998395}.

\subsection{Orthogonal Case}
In the case when the kinematic vector field is orthogonal to the fluid flow and if the similarity index $\delta \neq 0$, Eq.~\eqref{eq::kinematic_vector_field} becomes
\begin{equation} \label{eq::orthogonal_kssvfs}
    \boldsymbol{\xi}(t,r) = \xi^{\mu} \dfrac{\partial}{\partial x^{\mu}} = r \dfrac{\partial}{\partial r},
\end{equation}
and the self-similar variable $\xi$ becomes $t$. Similarly, we can define the line element in the co-moving frame 
\begin{equation}
    \mathrm{d}s^2 = - r^{2\alpha}\mathrm{d}t^2+ \mathrm{e}^{2\Psi(\xi)} \mathrm{d}r^2 + r^2\cR^2(\xi) \mathrm{d}\Omega^2.
\end{equation}
where the $\delta$ self-similarity index is not zero. If $\delta=0$, the KSSVF takes the form of $\boldsymbol{\xi} = \partial/\partial t$ and the metric can be expressed accordingly, 
\begin{equation} \label{eq::infty_orthogonal}
    \mathrm{d}s^2=-\mathrm{e}^{2r} \mathrm{d}t^2 + \mathrm{e}^{2\Psi(\xi)}\mathrm{d}r^2 + \cR^2(\xi) \mathrm{d}\Omega^2. 
\end{equation}
\subsubsection{Self-similarity of the First, Second, and Zeroth Kind}
The form of Einstein's field equations, as well as in the tilted solutions, implies the form of density, effective pressure, and MSH-mass, which will result in 
\begin{align}
    \kappa \rho & = r^{-2}\rho_0(\xi)+r^{-2\alpha} \rho_2(\xi),\\
    \kappa P & = r^{-2}P_0(\xi)+r^{-2\alpha} P_2(\xi),\\ 
    2m & = rm_0(\xi)+r^{3-2\alpha} m_2(\xi).
\end{align}
Since the EFEs are independently satisfied at both orders $\mathcal{O}(r^0)$ and $\mathcal{O}(r^{2-2\alpha})$. The relevant energy conservation equations will be
\begin{align}
    -\rho_0' & = (\rho_0+P_0)(\Psi'+2\cR'\cR^{-1}), \\
    -\rho_2' & = (\rho_2+P_2)(\Psi'+2\cR'\cR^{-1}), \\
    \alpha \rho_0 & = (2-\alpha)P_0,\\
    \alpha \rho_2 & = \alpha P_2. 
\end{align}
Similarly to the parallel case, the conservation equation implies that the $w(\xi)$ effective EoS parameter should be constant; hence, the bulk viscosity should trend to zero asymptotically.
\subsubsection{Self-similarity of Infinite Kind}
The form of the matter field is implied by the relevant dynamical equations, such as
\begin{align}
    \kappa \rho & = \mathrm{e}^{-2r}\rho_0(\xi)+\rho_2(\xi),\\
    \kappa P & = \mathrm{e}^{-2r}P_0(\xi)+ P_2(\xi),\\ 
    m & = \mathrm{e}^{-2r}m_0(\xi)+ m_2(\xi),
\end{align}
where the $\xi = t$ similarly to the previous cases and the metric is determined in Eq.~\eqref{eq::infty_orthogonal}. In this case, there is no consistent solution for the governing equation as analyzed by Hideki Maeda et al. \cite{HidekiHarada}.
\section{Discussion}
We provided a detailed analysis of how the kinematic self-similarity should be applied to Landau\,--\,Lifschitz\,--\,Eckart formalism. We have shown that the kinematic conditions impose very strict constraints on the form of the bulk viscosity, limiting the dependence of the bulk viscosity to $\zeta(\rho) \propto \rho^{1/2}$ and $\zeta(\theta) \propto \theta$. Our study reveals that the mathematical conditions (provided by the KSSVF framework) are often too restrictive, inevitably leading to either vacuum solutions or perfect-fluid solutions. Thus, we proved that if the KSSVF is either orthogonal or parallel to the fluid flow, it does not lead to solutions that involve viscous flows.
\begin{figure}[htbp]
    \centering
    \includegraphics[width=0.9\columnwidth]{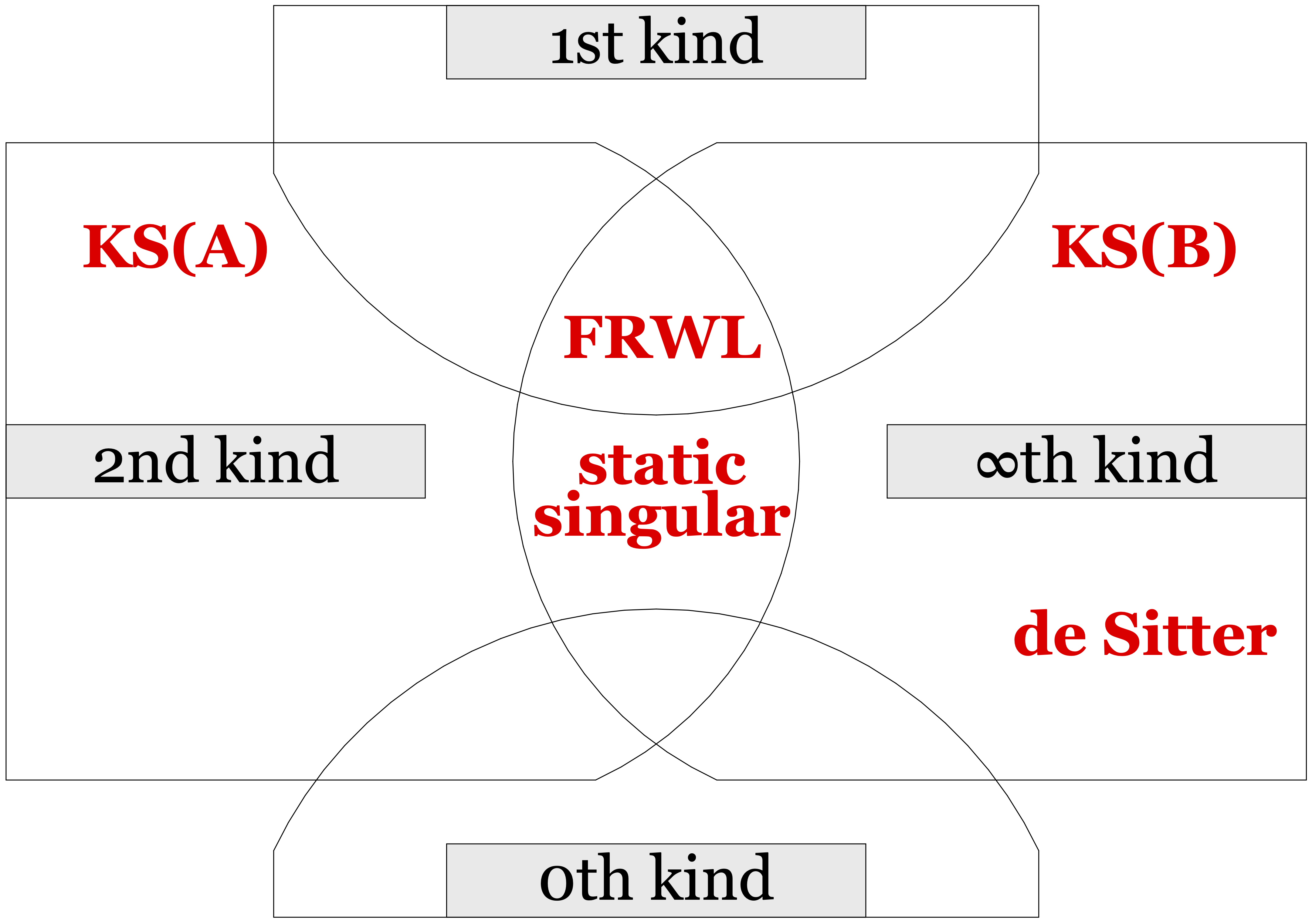} 
    \caption{The figure shows the possible 'tilted' kinematic self-similar solutions for spherically symmetric bulk-viscous space-times, classified according to the known geometries. KS(A) and KS(B) indicate two types of Kantowski\,--\,Sach solutions.}
    \label{fig::bulk_viscous_solution}
\end{figure}
In contrast to the classes discussed above, the extension to bulk viscous flows can be consistently carried out in the case of 'tilted' solutions. Their classification, along with the various possible solutions, is shown in Figure~\ref{fig::bulk_viscous_solution}. These results are in complete agreement with previous studies conducted on perfect fluid cases~\cite{Benoit1998, Henriksen1989, HidekiHarada}. The obtained solutions are identified with well-known cosmological space-times, including the Friedmann\,–-\,Robertson\,–-\,Walker\,–-\,Lemaître models~\cite{Friedmann1922}, the Einstein\,--\,de Sitter universe~\cite{1932PNAS...18..213E}, and the Kantowski–Sachs solutions~\cite{10.1063/1.1704952}. Building on the well-established similarity hypothesis introduced by Carr, which posits that spherically symmetric flows tend toward self-similar configurations at late times, our analysis of viscous models underscores the utility of self-similar solutions in clarifying their asymptotic properties. However, when the ultimate (large-scale) boundary conditions influence the evolution, the notion of self-similarity as a true asymptotic state must be modified, and the resulting solutions should rather be regarded as intermediate asymptotes in the sense of Barenblatt~\cite{Barenblatt1996, henriksen2015scale}.

A comparison of the asymptotic dynamics of viscous systems with those of perfect fluids indicates that the two systems display analogous behavior. In astrophysics, this implies that viscous effects may be challenging to constrain theoretically, since, in the context of kinematic self-similarity, both types of flows tend to similar space-time configurations at late times, making it difficult to distinguish the influence of viscosity solely on the asymptotic evolution of the system. This phenomenon is particularly relevant in cosmological contexts, during gravitational collapse~\cite{brandt}, and in the formation of compact objects~\cite{10.1093/mnras/stw2796}, where the long-term behavior of the fluid plays a crucial role. This trend is further supported by studies that find asymptotically perfect, spherically symmetric solutions of relativistic dissipative hydrodynamics in flat space using the Israel–Stewart theory, demonstrating that dissipation may not alter the late-time evolution qualitatively~\cite{Kasza}.
\section{Outlook}
A natural extension of this analysis is to move beyond the first-order Eckart\,--\,Landau\,--\,Lifshitz flow by adopting a second-order theory of relativistic dissipation, such as Müller\,--\,Israel\,--\,Stewart theory, which overcomes the well-known causality and instability issues of first-order theories. Extending the current classification scheme into this causal regime could uncover new classes of physically admissible solutions relevant to cosmology~\cite{PhysRevD.91.043532}, gravitational collapse, and the formation of compact objects. Similarly, analyzing shear-driven anisotropic stresses within the same frame would complete the description of viscous effects, offering a unified view of relativistic dissipation in strongly gravitating systems.
\section*{Acknowledgement}
We would like to express our heartfelt gratitude to István Szapudi, Norbert Barankai and Gábor Kasza for their insightful discussions and valuable support throughout this work. Their contributions and expertise have significantly enriched the work presented here. We also wish to acknowledge the Wigner Scientific Computing Laboratory for providing the necessary resources and support for the Wolfram Mathematica calculations. This work was supported by the research grants of No. NKFIH NKKP ADVANCED\_25-153456, NEMZ\_KI-2022-00058, 2025-1.1.5-NEMZ\_KI-2025-00005, 2024-1.2.5-TET-2024-00022 and FuSe COST Action CA-24101 for which we are deeply grateful.

\appendix
\section{Notation and Conventions} \label{Sec::Append1}

Our paper used the $c =1$, $h = 1$, $G = 1$ unit system and $\kappa = 8 \pi$. For arbitrary contravariant coordinates $x^{\mu}$, where Greek letters run from 0 to 3. We have used $g_{\mu \nu}$ covariant metric tensor with $(-,+,+,+)$ signature to describe the $4$-dimensional spacetime. The semicolon $X_{\mu;\nu}$ denotes the covariant derivative. Also, the prime $(')$ and $(\Dot{})$ represent the partial derivative by spatial and time coordinates, respectively. The $\mathrm{d}\Omega^2 = \mathrm{d}\theta^2 + \Sigma^2 (k,\theta)\mathrm{d}\phi^2$ 
\section{Four-velocity decomposition} \label{Sec::Append2}

The decomposition of the covariant derivative of the four-velocity is 
\begin{equation} \label{eq:ucomp}
    u_{\mu;\nu} = h^{\rho}_{\mu}h^{\kappa}_{\nu} u_{\rho;\kappa} + \Dot{u}_{\mu} u_{\nu} = b_{\mu \nu} + \Dot{u}_{\mu} u_{\nu},
\end{equation}
using the definition of the relevant covariant derivative \cite{Misner:1973prb}. The $b_{\mu \nu}$ can be decompose into symmetric and skew-symmetric part
\begin{equation}
    b_{\mu \nu} = \sigma_{\mu \nu} + \omega_{\mu \nu}.
\end{equation}
where,
\begin{equation}
    \sigma_{\mu \nu} = \theta_{(\mu \nu)} = h_{(\mu}^{\rho} h^{\kappa}_{\nu )} u_{\rho; \kappa}, \\
\end{equation}
and
\begin{equation}
    \omega_{\mu \nu} = \omega_{[\mu \nu]} = h_{[\mu}^{\rho} h^{\kappa}_{\nu ]} u_{\rho; \kappa}.
\end{equation}
The vorticity tensor, $\omega_{\mu \nu}$, is the skew-symmetric part. It determines the rigid rotation of the flow with respect to a local inertial rest frame. The symmetric part $\theta_{\mu \nu}$ is traditionally decomposed further into traceless and trace parts: 
\begin{equation}
    \theta_{\mu \nu} = \sigma_{\mu \nu} + \dfrac{1}{3} \theta h_{\mu \nu} \ ,
\end{equation}
so that 
\begin{equation}
    \theta^{\mu}_{\mu} = 0 \quad \text{and} \quad \theta = u^{\mu}_{;\mu} \ .
\end{equation}
We are also going to introduce the magnitude of the vorticity and the symmetric part:
\begin{equation}
    \sigma^2 =  \dfrac{1}{2} \sigma_{\mu \nu} \sigma^{\mu \nu}   \quad \text{and} \quad \omega^2 = \dfrac{1}{2} \omega_{\mu \nu} \omega^{\mu \nu}, 
\end{equation}
Finally, the acceleration vector is defined in Eq.~\eqref{eq:ucomp} in the following way
\begin{equation}
    \Dot{u}^{\mu} = u^{\mu}_{; \nu} u^{\nu}. 
\end{equation}

\section{Lie-derivative of the pressure, four-velocity and energy density in the case of viscous energy tensor} \label{sec::AppendixLie}

Let $u^{\mu}$ be a timeline vector field normalized in such way $u_{\mu}u^{\mu} = -1$. The Lie-derivative $\LieXi$ along a homothetic vector field $\boldsymbol{\xi}$ can be expressed in the following way, 
\begin{align}
    0 &= \mathcal{L}_{\xi} \left(u_{\mu}u^{\mu} \right) =  \mathcal{L}_{\xi} \left(g_{\mu \nu}u^{\mu}u^{\nu} \right) \notag \\
    & = 2 u_{\mu} \mathcal{L}_{\xi} u^{\mu} + u^{\mu} u^{\nu} \mathcal{L}_{\xi} g_{\mu \nu} = 2 u_{\mu} \mathcal{L}_{\xi} u^{\mu} - 2 \alpha.
\end{align}
Consequently, the Lie derivative of the covariant and contravariant timeline vector field $u_{\mu}$ should be
\begin{equation}
    \mathcal{L}_{\xi} u^{\mu} = - u^{\mu} - v^{\mu} \quad \text{and} \quad \mathcal{L}_{\xi} u_{\mu} = u_{\mu} - v_{\mu}.  
\end{equation}
where $u_{\mu}$ and $v_{\mu}$ are orthogonal to each other. According to the Eq.~\eqref{eq::LieDerivativeT} and the form of the viscous energy-momentum tensor Eq.~\eqref{eq:ViscousEnergyTensor} we can derive the following relation,
\begin{align}
    0 & = \mathcal{L}_{\xi} T_{\mu \nu} \notag \\
      & = \mathcal{L}_{\xi}\left[(\rho + P) u_{\mu} u_{\nu} + P g_{\mu \nu} + 2 \eta \sigma_{\mu \nu} \right] \notag \\
      & = (\mathcal{L}_{\xi}\rho + \mathcal{L}_{\xi}P)u_{\mu}u_{\nu} + (\mathcal{L}_{\xi}P)g_{\mu\nu} \notag \\
      & \ \ + (\rho+P)\mathcal{L}_{\xi}(u_{\mu}u_{\nu}) + P\mathcal{L}_{\xi}g_{\mu \nu} + 2\eta\mathcal{L}_{\xi}\sigma_{\mu \nu} \notag \\
      & = (\mathcal{L}_{\xi}\rho + \mathcal{L}_{\xi}p)u_{\mu}u_{\nu} + (\mathcal{L}_{\xi}p)g_{\mu\nu} \notag \\
      & \ \ + (\rho + p)\mathcal{L}_{\xi}(u_{\mu}u_{\nu}) + p\mathcal{L}_{\xi}g_{\mu \nu} + 2\eta\mathcal{L}_{\xi}\sigma_{\mu \nu} \notag \\
      & \ \ + \LieXi \Pi (g_{\mu \nu} + u_{\mu} u_{\nu}) + \Pi \LieXi g_{\mu \nu}, 
\end{align}
if we use the formula for the projection tensor $h_{\mu \nu}$, then 
\begin{align} \label{eq::energy-stress-lie}
    0 & = (\mathcal{L}_{\xi} \rho + 2 \alpha \rho)u_{\mu}u_{\nu} + \left(\mathcal{L}_{\xi}  + 2 \alpha \right)\left[ p+ \Pi \right]h_{\mu\nu}  \notag \\
     & \ \ - (\rho + p + \Pi) (u_{\mu} v_{\nu} - u_{\nu} v_{\mu}) + 2 \eta \LieXi \sigma_{\mu \nu}. 
\end{align}
For arbitrary vector field $V^{\mu}$ and $\xi$ homothetic vector field the following identity holds
\begin{equation}
\mathcal{L}_\xi (\nabla_\mu V^\mu) \overset{!}{=} \nabla_\mu (\mathcal{L}_\xi V^\mu).
\end{equation}
The general commutation relation is 
\begin{align}
& [\nabla_{\mu},\LieXi]V^{\mu} = \nabla_\mu (\mathcal{L}_\xi V^\mu) -  \mathcal{L}_\xi (\nabla_\mu V^\mu) = \\ \notag
& \ \ = (\nabla_\mu \xi^\nu)(\nabla_\nu V^\mu)  - (\nabla_\mu V^\nu)(\nabla_\nu \xi^\mu) 
- V^\nu \nabla_\mu \nabla_\nu \xi^\mu.
\end{align}
Applying the decomposition of $\nabla_{\mu}\xi^{\nu} = 2\alpha \delta^{\nu }_{\mu} + \omega_{\mu}^{\nu}$ where $\omega_{\mu \nu}$ is antisymmetric, it becomes trivial to show that the first and seconds terms on the r.h.s will vanish. Using the Ricci identity, the third term transforms as, 
\begin{equation}
\nabla_\mu \nabla_\nu \xi^\mu = - R_{\mu\nu} \xi^\mu \Rightarrow - V^\nu \nabla_\mu \nabla_\nu \xi^\mu =   V^\nu R_{\mu\nu} \xi^\mu,
\end{equation}
and it will vanish if $\xi$ is a proper HVF \cite{yano1957lie,Stephani2003}. By contracting Eq.~\eqref{eq::energy-stress-lie} in different ways and set $\sigma_{\mu \nu}$ to zero, it follows that
\begin{align}
    u^{\mu} u^{\nu}\mathcal{L}_{\xi} T_{\mu \nu} = 0  &  \rightarrow \mathcal{L}_{\xi} \rho = - 2 \alpha \rho, \\
    h^{\mu}_{\sigma} h^{\nu}_{\rho} \mathcal{L}_{\xi} T_{\mu \nu} =  0  & 
 \rightarrow \mathcal{L}_{\xi} P = - 2 \alpha P,  \\
     u^{\mu} h^{\nu}_{\rho} \mathcal{L}_{\xi} T_{\mu \nu}   =  0  &  \rightarrow  (\rho + P) v_{\sigma} = 0.  \label{eq::Lie_Derivative_const}
\end{align}
The last relation of Eq.~\eqref{eq::Lie_Derivative_const} implies that if $w(\xi) \neq -1$ and non-vacuum solution is considered, the $v_{\sigma}$ must vanish, then
\begin{equation}
    \LieXi u^{\mu} = - u^{\mu} \quad \text{and} \quad \LieXi u_{\mu} = u_{\mu} \ .
\end{equation}
\section{General Spherically Symmetric Spacetime} \label{sec::appendB}
From the line element Eq.~\eqref{eq::line_element} of the most general spherically symmetric space-time, the non-zero elements of the Christoffel symbol are
\begin{align}
\Gamma^{t}_{t t} &= \dot{\phi}, \qquad \Gamma^{t}_{t r} = \Gamma^{t}{}_{r t} = \phi' \notag\\
\Gamma^{t}_{r r} &= \dot{\psi} \mathrm{e}^{2 \psi - 2 \phi}, \Gamma^{t}_{\varphi \varphi} = \Sigma(k, \theta)^2 \Gamma^{t}_{\theta \theta}, \quad  \Gamma^{t}_{\theta \theta} = \textrm{e}^{-2 \phi} R  \dot{R}, \notag \\
\Gamma^{r}_{t t} & = \textrm{e}^{2 \phi - 2 \psi} \phi', \quad \Gamma^{r}_{t r} = \Gamma^{r}{}_{r t} = \dot{\psi}, \notag \\
\Gamma^{r}_{r r} &= \psi', \quad \Gamma^{r}_{\theta \theta} =  -\textrm{e}^{-2 \psi} R R', \quad \Gamma^{r}_{\varphi \varphi} = \Sigma^2(k, \theta) \Gamma^{r}_{\theta \theta} , \notag \\
\Gamma^{\theta}_{t \theta} &= \Gamma^{\theta}{}_{\theta t} = \Gamma^{\varphi}{}_{t \varphi} = \Gamma^{\varphi}_{\varphi t} =\dot{R}R^{-1}, \notag \\
\Gamma^{\theta}_{r \theta} &= \Gamma^{\theta}{}_{\theta r} = \Gamma^{\varphi}{}_{r \varphi} = \Gamma^{\varphi}_{\varphi r} = R'R^{-1}, \notag \\
\Gamma^{\theta}_{\varphi \varphi} &= \Sigma(k,\theta) \Sigma'(k,\theta), \quad \Gamma^{\varphi}_{\theta \varphi} = \Gamma^{\varphi}_{\varphi \theta} = \Sigma^{-1}(k,\theta) 
\end{align}
The non-zero components of the Ricci-tensor are, 
\begin{align}
    R_{t}^{t} & = R^{-1}\bigg\{\textrm{e}^{-2\Phi }\big(R(\Ddot{\Psi} +\Ddot{\Psi}^2-\Dot{\Phi} \Dot{\Psi}) + 2\Ddot{R} -2\Dot{R}\Dot{\Phi}\big) \notag \\
    & \ \ -\textrm{e}^{-2\Psi}\big(R\Phi'' + \Phi'\left(R (\Phi'-\Psi')+2R'\right)\big)\bigg\},\\
    R_{t}^{r}&=2\textrm{e}^{-2\Psi }R^{-1}\big(-\Dot{R}'+R'\Dot{\Psi}+\Dot{R}\Phi'\big),\\
    R_{t}^{r}&=2\textrm{e}^{-2\Phi }R^{-1}\big(-\Dot{R}'+R'\Dot{\Psi}+\Dot{R}\Phi'\big),\\
    R_{r}^r & = \textrm{e}^{-2\Psi}\bigg[ R^{-1} \bigg( \textrm{e}^{2(\Psi-\Phi)}\left(R\Ddot{\Psi}+\Dot{\Psi}\left(R\left(\Dot{\Psi} -\Dot{\Phi}\right)+2\Dot{R}\right)\right)  \notag \\
    & \ \ -R\Phi''-2R''+2R'\Psi' \bigg)+\Phi'\left(\Psi'-\Phi'\right)\bigg],\\
    R_{\theta}^{\theta} &  = \frac{\textrm{e}^{-2(\Psi +\Phi)}}{R^2} \bigg[ \textrm{e}^{2 \Psi} \big( \textrm{e}^{2 \Phi} + R \Ddot{R} + R \Dot{R} (\Dot{\Psi} - \Dot{\Phi}) + \Dot{R}^2 \big) \notag \\
    & \ \ - \mathrm{e}^{2\Phi} \big( R R'' + R R' (\Dot{\Phi} - \Dot{\Psi}) + (R')^2 \big)\bigg]\\
    R_{\phi}^{\phi} & = \Sigma^2(k,\theta)  R_{\theta}^{\theta},
 \end{align}%
By taking the trace of the Ricci tensor with respect to the metric tensor, the Ricci scalar is obtained, such as
\begin{align}
R = &\frac{\mathrm{e}^{-2 (\Psi + \Phi)}}{R^2} \Big\{ 
2 \textrm{e}^{2 \Psi} \Big[ \textrm{e}^{2 \Phi} + 2 R \dot{R} (\dot{\Psi} - \dot{\Phi}) + \dot{R}^2 + 2 R \ddot{R} \nonumber \\
& + R^2 (\dot{\Psi}^2 + \ddot{\Psi} - \dot{\Psi} \dot{\Phi}) \Big)  - 2 \textrm{e}^{2 \Phi} \Big( 2 R R' (\Phi' - \Psi') \nonumber \\ 
& + R \big( 2 R'' + R (-\Psi' \Phi' + \Phi'^2 + \Phi'') \big) + R'^2 \Big] \Big\}. 
\end{align}

\section{In-Depth Analysis and Derivation of Tilted Self-similar Solutions} \label{sec::appendixE}
\subsection{Self-similarity of Second-Kind}
First we eliminated the second derivative $\mathcal{R}''$ using Eq.~\eqref{eq::2ndrt} and we find,
\begin{equation}
    \mathcal{R}'' = \mathcal{R}'\Phi' + (\mathcal{R} + \mathcal{R}')\Psi' - \mathcal{R}'. \label{eq:Rdoubleprime}
\end{equation}
Substituting this formula into Eq.~\eqref{eq::aux_2nd_1}, followed by simplification, the auxiliary equation yields
\begin{equation}
    2\mathcal{R}\mathcal{R}'\Phi' + 2\mathcal{R}\mathcal{R}' = -\mathcal{R}'^2 - \mathcal{R}^2 + \textrm{e}^{2\Psi}(1 - \rho_0\mathcal{R}^2). \label{eq:simp_aux_1}
\end{equation}
Subtracting Eq.~\eqref{eq:simp_aux_1} from Eq.~\eqref{eq::aux_2nd_2} we arrive at the following equation
\begin{align}
    &\left[\mathcal{R}^2 + \mathcal{R}'^2 + 2\Phi'\mathcal{R}(\mathcal{R} + \mathcal{R}')\right] - \left[2\mathcal{R}\mathcal{R}'\Phi' \right] \nonumber \\
    &= \left(1 + P_0 \mathcal{R}^2\right)\textrm{e}^{2\Psi} + \left[\mathcal{R}'^2 \mathcal{R}^2 - \textrm{e}^{2\Psi}(1 - \rho_0\mathcal{R}^2)\right].
\end{align}
Canceling the common terms on both sides, we finally obtain a constraint equation between the pressure, the density and metric functions
\begin{equation}
    2\Phi' = ( \rho_0 +P_0)\textrm{e}^{2\Psi}. \label{eq:phi_constraint}
\end{equation}
From the relation between Eq.~\eqref{eq::EquationOrder1} and \eqref{eq::EquationOrder2} the effective pressure is assumed to be $P_0 = w(\xi) \rho_0$, and $P_2 = w(\xi) \rho_2$. Substituting these into Eqs.~\eqref{eq::2ndEq1}-\eqref{eq::2ndEq2} will result in
\begin{align}
    (1 + w)\Phi'  
    &= \left(2 - \frac{\rho_0'}{\rho_0} \right) w - w',  \\
    (1 + w)\Phi' &= - \left( w \frac{\rho_2'}{\rho_2} + w' \right).  
\end{align}
By forming the ratio of these equations, one obtains
\begin{align} 
\dfrac{\rho_0'}{\rho_0} = 2 + \dfrac{\rho'_2}{\rho_2}.
\end{align}
Similarly the Eqs.~\eqref{eq::2ndEq3}-\eqref{eq::2ndEq4} will lead to the relation of
\begin{align}
\dfrac{\rho_0'}{\rho_0} = 2\alpha + \dfrac{\rho'_2}{\rho_2}.
\end{align}
This leads us that the $\alpha = 1$, if $\rho_0\rho_2 \neq 0$. Concludes that the self-similar solution of the second kind can exist only if $\rho_0\rho_2 = 0$, similarly to the perfect fluid case \cite{HidekiHarada}.

\textbf{The $\boldsymbol{\rho_0 = 0}$ case: }If we assume that $\rho_0$ and consequently $P_0 = 0$, from Eq.~\eqref{eq::2ndconds_1} and from Eq.~\eqref{eq::2ndm1} we obtained that $m_0 = 0$. Neglecting the trivial case, the Eq.~\eqref{eq::misner_2nd_1} leads to the constraint of
\begin{equation} \label{eq::r_phi_relation}
    (\mathcal{R} + \mathcal{R}') = \pm \textrm{e}^{\Psi}.
\end{equation}
Likewise, the Eq.~\eqref{eq:phi_constraint} implies that the metric function of $\Phi = \ln\phi_0$ is constant. If we substitute this into Eq.~\eqref{eq::2ndEq2}, it will result that $P_2$ is constant and 
\begin{equation} \label{eq::rho_0_constrain}
    w'\rho_2 + \rho_2'w= 0,  \qquad \Rightarrow \qquad w(\xi) \rho_2(\xi)= \xi_0.
\end{equation}
Combining the remaining equations of Eqs.~\eqref{eq::2ndm2}, \eqref{eq::2ndcons_2}, and \eqref{eq::misner_2nd_2}, one can eliminate the $m_2$ and $m_2'$ terms and obtain the non-linear evolution equation of
\begin{equation}
    \rho_2 = U_0 \dfrac{U(\xi)}{1+U(\xi)}, \quad \Rightarrow \quad U_0 := \left( \xi_0 + \dfrac{3-2\alpha}{\alpha^2 \Phi_0^2} \right).
\end{equation}
where $U(\xi) = \cR'/\cR$. Substituting the obtained expression for $\rho_2$ into Eq.~\eqref{eq::2ndEq4}, will result in
\begin{equation} \label{eq::2nd_rho0_evolution}
    \dfrac{2\alpha U}{1+U} + \dfrac{U'}{(1+U)^2} =\left[\dfrac{\xi_0}{U_0} - \dfrac{U}{1+U} \right](\Psi'+2U).
\end{equation}
It can be transformed into a Ricatti-type equation by substituting the following relation obtained from Eq.~\eqref{eq::r_phi_relation}
\begin{equation}
    \Psi' = U +\dfrac{U'}{1+U}
\end{equation}
The closed-form solution for $U(\xi)$ is
\begin{equation}
 U(\xi) = \dfrac{C_U}{3+(1-\tilde{U}_0)^{C_U-1}\xi^{-C_U}},
\end{equation}
where $\tilde{U}_0 = \xi_0/U_0$ , the $C_U = (3\tilde{U}_0-2\alpha)/(1-\tilde{U}_0)$ and the integration constant is neglected. Applying the definition of $U(\xi)$ and integrating over the self-similarity variable, one can express the $\cR(\xi)$ as a function of $\xi$.
\begin{equation}
    \cR = \cR_0 \left( 3(1-\tilde{U}_0) \xi^{C_U} + (1-\tilde{U}_0)^{C_U}\right)^{1/3} .
\end{equation}
Applying this formula in Eq.~\eqref{eq::r_phi_relation} one can solve the ODE for the $\Psi(\xi)$ function
\begin{align} \label{eq::solution_psi_2nd}
    \Psi(\xi) & = \ln \Bigg( \cR_0 \xi ^{C_U/3} \bigg( 3-2\alpha+(1-\tilde{U}_0)^{C_U}\xi^{-C_U} \bigg) \notag \\
    &\times \left( 3(1-\tilde{U_0}) + (1-\tilde{U_0})^{C_U}\xi^{-C_U}\right)^{-2/3}\Bigg),
\end{align}
where $\cR_0$ is some integration constant. The Eq.~\eqref{eq::r_phi_relation} yields a complex solution in addition to the Eq.~\eqref{eq::solution_psi_2nd}. However, this complex solution is typically discarded in physical applications. The expression of Eq.~\eqref{eq::rho_0_constrain} and the definition of the effective $w(\xi)$ reveal that there are only two possibilities: either $\zeta \to 0$, which reduces to the perfect fluid case, or if $w_0, U_0 \to 0$, corresponding to the vacuum case. However, the $\zeta\propto \theta$ bulk viscosity form is consistent with the obtained solution if $w_0$ vanishes, the $\xi_0/\zeta U_0$ is set to unity and the $(1-U_0)^3\overset{!}{=}9$. 

\textbf{The $\boldsymbol{\rho_2=0}$ case: }In the second case, where $\rho_0 \neq 0$, the $\rho_2$ and $P_2$ are zero. From the Eq.~\eqref{eq::2ndEq3} and Eq.~\eqref{eq::aux_2nd_0} one can obtain the equation of,
\begin{equation} \label{eq::constrain_on_2ndcase}
\dfrac{1}{1+w(\xi)} \dfrac{\rho_0'}{\rho_0} = - \dfrac{3}{2}\frac{\cR'}{\cR} 
\end{equation}
Since the $P_2=0$, the Eq.~\eqref{eq::aux_2nd_3} becomes more trivial,
\begin{equation} \label{eq::new_aux_2nd}
      \cR''  - \Phi' \cR' = - \frac{1}{2}\left(\dfrac{\cR'}{\cR} + 2\alpha \right)\cR'.
\end{equation}
Using Eq.~\eqref{eq::2ndrt} we can eliminate the $\cR'' - \Phi' \cR'$ from Eq.~\eqref{eq::new_aux_2nd}. We can obtain an expression for $\Psi'$ and substituting this formula to Eq.~\eqref{eq::2ndEq3}, we have
\begin{align} \label{eq::2nd_aux_2_UV}
    (\alpha-1)U(\xi) + \dfrac{1}{2}U^2(\xi) & - \notag \\  \left[1+U(\xi)\right]\left( 2U(\xi) +\dfrac{1}{1+w(\xi)}V(\xi)\right)  & = 0,
\end{align}
where $V(\xi)=\rho_0'/\rho_0$. Applying the constrain from the Eq.~\eqref{eq::constrain_on_2ndcase}, we have
\begin{equation}
    \left(2\alpha - 9\right)U(\xi)-6U^2(\xi)=0.
\end{equation}
Solving this non-linear equation provides two distinct solutions. The trivial solution corresponds to the case where $U(\xi) = 0$ and $\cR$ remains fixed at the constant value $\cR_0$. The non-trivial solution is 
\begin{equation} \label{eq::non_trivial_solution}
    U(\xi) = \dfrac{1}{3}\left(\alpha - \dfrac{9}{2}\right) \Rightarrow \cR(\xi) = \cR_0  \xi^{\lambda_{0,r}}, 
\end{equation}
where $\lambda_{0,r} = (2\alpha-9)/6$ is some constant.

\textit{(A.)} Applying the non-trivial solution the Eq.~\eqref{eq::new_aux_2nd} and the Eq.~\eqref{eq::aux_2nd_0} becomes
\begin{equation}
    \Phi(\xi) = c_{\Phi} \ln \xi \quad \text{and} \quad \Psi(\xi) = -\dfrac{1}{2}\lambda_{0,r} \ln \xi ,
\end{equation}
where $c_{\Phi} = 3(2\alpha-3)/4$. Hence, we can use the Eq.~\eqref{eq::constrain_on_2ndcase} to find $\rho_0$ if the bulk viscosity parameter depends on the density $\sim \rho^{1/2}$
\begin{align}
\rho_0(\xi)  & = \bigg( 
\frac{\tfrac{3}{4}\zeta \lambda_{0,r}^2}{2\left(-c_{\Phi} + \tfrac{3}{4}\lambda_{0,r}(1+w_0)\right)} \,\xi^{-c_{\Phi}} + \notag \\
& + \, c_{\rho,0} \,\xi^{-\tfrac{3}{4}\lambda_{0,r}(1+w_0)}
\bigg)^{2}
\end{align}
where $c_{\rho,0}$ is some integration constant. Finally, one can obtain the following formula for the $P_0$
\begin{align}
P_0(\xi) & = w_0 \bigg(
\frac{\tfrac{3}{4}\zeta \lambda_{0,r}^2}{2\left(-c_{\Phi} + \tfrac{3}{4}\lambda_{0,r}(1+w_0)\right)} \,\xi^{-c_0} \, + \notag \\
& + c_{\rho_0} \,\xi^{-\tfrac{3}{4}\lambda_{0,r}(1+w_0)}
\bigg)^{2} \notag \\ 
& - \tfrac{1}{2}\zeta \lambda_{0,r}\,\xi^{-c_0}\bigg(
\frac{\tfrac{3}{4}\zeta \lambda_{0,r}^2}{2\left(-c_{\Phi} + \tfrac{3}{4}\lambda_{0,r}(1+w_0)\right)} \,\xi^{-c_0} \notag \\ 
& + C_{\rho_0} \,\xi^{-\tfrac{3}{4}\lambda_{0,r}(1+w_0)}
\bigg).
\end{align}
Consequently from the definition of the MSH-mass Eqs.~\eqref{eq::misner_2nd_1}-\eqref{eq::misner_2nd_2} we obtain the formula for $m_0$ and $m_2$
\begin{equation}
m_0 = \cR_0\, \xi^{\lambda_{\rho,0}} \left[ 1 - (1 + \lambda_{\rho,0})^2 \cR_0^2 \, \xi^{4 \lambda_{\rho,0}} \right], 
\end{equation}
and the $m_2$ will vanish according to the Eq.~\eqref{eq::2ndm2} and Eq.~\eqref{eq::2ndcons_2}.

\textit{(B.)} Applying the trivial solution of $U(\xi)$, the radial term is $\cR(\xi) = \cR_0$.  Hence the $\Psi = \ln \psi_0$ and $m_0 = \cR_0(1-\cR_0^2/\psi_0^2)$ also became constant based on the obtained from Eqs.~\eqref{eq::2ndrt}-\eqref{eq::misner_2nd_1}. The Eq.~\eqref{eq::2ndm1} implies that $\rho_0$ must be constant
\begin{equation}
    \rho_0 = \dfrac{\cR^2_0 - \psi_0^2}{\cR^2_0 \psi_0^2}.
\end{equation}
It is apparent from the definition of the $w(\xi)$ that the bulk viscosity part will vanish and become a constant of $w_0$, and the $P_0$ will be a constant. It is evident from Eq.~\eqref{eq::2ndEq1} that $\Phi(\xi)$ can expressed in terms of $\xi$ and $w_0$ in the following way,
\begin{equation}
    \Phi(\xi) = \dfrac{2w_0}{(1+w_0)} \ln \xi
\end{equation}
\subsection{Self-Similar of Zeroth Kind} \label{sec::sec::zeroth}

Since the governing equations do not change drastically if we impose $0^{th}$ kind of self-similarity, the derived constraint for $\Phi$ described in Eq.~\eqref{eq:phi_constraint} still holds. Also, the obtained condition of $\rho_0(\xi)\rho_2(\xi)=0$, required for the existence of a solution, remains valid.

First, we require the $\rho_0$ to be zero, and since $P_0$ and also $m_0$ vanish trivially, applying the EoS and the Eq.~\eqref{eq::0th_m0_def}. Hence, the Eq.~\eqref{eq:phi_constraint} suggest that $\Phi=\ln\phi_0$ to be constant. Since $m_0=0$, the definition of MSH mass indicates that 
\begin{equation} \label{eq::0th_psi_constraint}
    (\cR+\cR') = \pm\mathrm{e}^{\Psi},
\end{equation}
if we neglect the solution, where $\cR(\xi)=0$. One can express the $\Psi(\xi)$ variable and substitute into Eq.~\eqref{eq::2ndrt} will result in,
\begin{equation}
    \cR(\xi) = \cR_0\xi \quad \text{and} \quad \Psi(\xi)=\ln(2\cR_0\xi).
\end{equation}
The minus sign in Eq.~\eqref{eq::0th_psi_constraint} would correspond to $\cR_0 < 0$,  which leads to a complex solution and therefore it is nonphysical, thus disregarded. Consequently, the definition of second-order MSH mass of Eq.~\eqref{eq::misner_2nd_2} (with $\alpha=1$) implies that $m_2(\xi) = \cR_0^3\Phi_0^2\xi^3$. Also, the Eq.~\eqref{eq::0th_m2_def} suggest that the pressure equals $P_2 = -3\Phi_0^2$. That indicates that $\rho_2$ is also constant and the solution is equivalent to the perfect fluid case. Also, the Eq.~\eqref{eq::0th_m2_def} suggest that the pressure equals $P_2 = -3\Phi_0^2$. That indicates that $\rho_2$ is also constant and the solution is equivalent to the perfect fluid case. 

Secondly, we have the case where $\rho_2 = 0$ and consequently $P_2$ vanish. The Eq.~\eqref{eq::misner_2nd_2} and Eq.~\eqref{eq::2ndm2} suggest that $m_2$ should also vanish. If we take the $0^{th}$ self-similar form of the Eq.~\eqref{eq::new_aux_2nd}, it becomes
\begin{equation}
  \cR'' - \Phi'\cR' = - \dfrac{1}{2} \dfrac{\cR'^2}{\cR}
\end{equation}
and similarly Eq.~\eqref{eq::2nd_aux_2_UV} transforms as 
\begin{equation} \label{eq::0th_eq_1}
    3U(\xi) + \dfrac{3}{2}U^2(\xi) + \bigg( 1+U(\xi) \bigg)\left(\dfrac{1}{1+w(\xi)}V(\xi)\right) = 0,
\end{equation}
where $U(\xi) = \cR'/\cR$ and $V(\xi)=\rho_0'/\rho_0$ respectively. Also, similarity to the $2^{nd}$ kind of self-similarity case we have the 
\begin{equation} \label{eq::0th_eq_2}
\dfrac{1}{1+w(\xi)} V(\xi)= - \dfrac{3}{2} U(\xi).
\end{equation}
It is trivial to see from Eq.~\eqref{eq::0th_eq_1} and Eq.~\eqref{eq::0th_eq_2}, which lead to $U(\xi)$ being zero and $\cR(\xi) = c_{\cR}$. Consequently, $\rho_0 = c_{\rho,0}$ will be constant. Substituting the obtained solutions into Eq.~\eqref{eq::0th_m0_def} will result in $m_0(\xi)$ being constant. Also the $\Psi$ can be determined from MSH mass definition of Eq.~\eqref{eq::misner_2nd_1}
\begin{equation} \label{eq::m0_0th_res}
    \Psi(\xi) = - \dfrac{1}{2} \ln \left( \dfrac{c_{\cR} - c_{m}}{c_{\cR}^2}\right).
\end{equation}
Note that these constants are not fully independent, such Eq.~\eqref{eq::misner_2nd_1} implies that $c_m \overset{!}{=} - c_{\rho,0}c_{\cR}^3$. The last equation of Eq.~\eqref{eq::2ndEq1} reveals that $w(\xi)$ must be constant, and the viscosity term will vanish. The obtained $\Phi(\xi) $ metric function is 
\begin{equation}
    \Phi(\xi) = \dfrac{2w_0}{w_0+1} \ln \xi + c_{\Phi}.
\end{equation}
\subsection{Self-similarity of Infinite Kind} \label{sec::sec::infinite_tilted}
The form of the density and effective pressure in Eq.~\eqref{eq::infti_dens}-\eqref{eq::infti_press} imply that the $P_0=w(\xi)\rho_0$ and $P_2=w(\xi)\rho_2$ similarly to the previous cases. First, we can express the $\cR''$ and $\Psi'$ from Eq.~\eqref{eq::rt_equation_infinite} and Eq.\eqref{eq::infinity_aux_3} and substitute into the Eq.~\eqref{eq::infinity_aux_4} and we arrive the result of,
\begin{equation} \label{eq::infinity_cons}
    \bigg(1+w(\xi)\bigg)\rho_0\cR^2\mathrm{e}^{2\Psi} = 0
\end{equation}
The energy conservation Eqs.~\eqref{eq::infinity_rho_0}-\eqref{eq::infinity_rho_2} imply that $\rho_0(\xi) =  C\rho_2(\xi)$ and consequently from Eq.~\eqref{eq::infinity_cons} we have
\begin{equation}
    w(\xi) = -1 \quad \text{or} \quad \rho_0 =0 \Rightarrow C\overset{!}{=}0,
\end{equation}
since we are not interested in vacuum solutions. One can notice that the condition $w(\xi)=-1$ reduces the problem to the perfect fluid solution, and hence, no bulk viscous solution exists. If $\rho_0=0$, then trivially from Eq.~\eqref{eq::infinity_mass_cons_0} the mass $m_0$ is constant and chosen to be zero. Let's assume that $\mathrm{e}^{\Phi} = \Phi_0$ than from Eq.~\eqref{eq::infinity_rho_2} we obtain that 
\begin{equation}
    P_2 = c_{P_2} \quad \textrm{and} \quad w(\xi)\rho_2(\xi) = c_{P_2}.
\end{equation}
We repeat the process from the first step and express the $\cR''$ and $\Psi'$ from Eqs.~\eqref{eq::rt_equation_infinite}-\eqref{eq::infinity_aux_1} and substitute into Eq.~\eqref{eq::infity_aux_2} and
\begin{equation} \label{eq::rho2_expression}
    \rho_2 = -2\Phi^{-2}_{0}\dfrac{\cR'}{\cR}.
\end{equation} 
 From the definition of the MSH mass, we have $\cR' = \pm \mathrm{e}^{\Psi}$ and substituting it and $\rho_2$ from Eq.~\eqref{eq::rho2_expression} into Eq.~\eqref{eq::infinity_eq_4} we have obtained
 \begin{equation} \label{eq::infinity_evolution}
    -4\Phi_0^{-2}U-2\Phi_0^{-2}(U'+U^2) = (2\Phi_0^{-2}U + c_{P_2})\left( 3U + \dfrac{U'}{U}\right)
 \end{equation}
 where $U= \cR'/\cR$. This differential equation cannot be solved by analytical terms. However, we could assume that $U'$ vanishes and the ODE reduces to an algebraic equation of
 \begin{equation}
     U(4+3c_{P_2}\Phi_0^2 + 8U)=0,
 \end{equation}
which yield two constant solutions
\begin{equation}
    U_1 = 0 \quad \mathrm{and} \quad U_2= -\dfrac{1}{8}(4+ 3 c_{P_2}\Phi_0^2).
\end{equation}
To assess the stability of the constant solutions to the differential equation when 
$U'$ is infinitesimally small, we perform a linear stability analysis $U(\ln \xi) = U_0 + \delta(\ln \xi)$ around each $U_0$ fixed point. One can clearly see, that the term of $U'/U$ from Eq.~\eqref{eq::infinity_evolution} diverges at $U_1=0$ which makes it a singular point and consequently unstable. However, the linear stability analysis indicates that $U_2$ fixed point is regular and generally stable if $c_{P_2} > - 4/3\Phi_0^{-2}$. Applying the definition of $U$, we obtain the following expressions for $\cR$ and $\Psi$
\begin{equation}
    \cR(\xi) = \cR_0 \xi^{U_2} \quad \mathrm{and} \quad  \mathrm{e}^{\Psi} = \pm U_2\cR_0\xi^{U_2}.
\end{equation}

\bibliography{apssamp}

\end{document}